\documentclass[12pt,preprint]{aastex}

\shorttitle{Star formation in NGC~346}
\shortauthors{Sabbi et al.}

\begin{document}

\title{Past and present star formation in the SMC: NGC~346 and its
neighborhood\footnote{Based on observations with the NASA/ESA Hubble Space
Telescope, obtained at the Space Telescope Science Institute, which is
operated by AURA Inc., under NASA contract NAS 5-26555}}

\author{E.~Sabbi\altaffilmark{2}, M.~Sirianni\altaffilmark{2,3}, 
A.~Nota\altaffilmark{2,3}, M.~Tosi\altaffilmark{4},  
J.~Gallagher\altaffilmark{5}, M.~Meixner\altaffilmark{2}, M.~S.\ 
Oey\altaffilmark{6}, R.~Walterbos\altaffilmark{7}, A.~Pasquali\altaffilmark{8}, 
L.~J.\ Smith\altaffilmark{9} \& L.~Angeretti\altaffilmark{4}}
\email{sabbi@stsci.edu}

\altaffiltext{2}{Space Telescope Science Institute, 3700 San Martin Drive,
Baltimore, USA}
\altaffiltext{3}{ESA, Space Telescope Operation Division}
\altaffiltext{4}{INAF--Osservatorio di Bologna, I}
\altaffiltext{5}{University of Wisconsin, USA}
\altaffiltext{6}{University of Michigan, USA}
\altaffiltext{7}{University of New Mexico State, USA}
\altaffiltext{8}{MPIA, D}
\altaffiltext{9}{University College London, UK}

\begin{abstract}
In the quest of understanding how star formation occurs and propagates in the
low metallicity environment of the Small Magellanic Cloud (SMC), we acquired
deep F555W ($\sim$V), and F814W ($\sim$I) HST/ACS images of the young and
massive star forming region NGC 346. These images and their photometric
analysis provide us with a snapshot of the star formation history of the
region. We find evidence for star formation extending from $\approx$10~Gyr in
the past until $\approx 150\, {\rm Myr}$ in the field of the SMC. The youngest stellar
population ($\sim 3\pm 1\, {\rm Myr}$) is associated with the NGC~346 cluster. It
includes a rich component of low mass pre-main sequence stars mainly
concentrated in a number of sub--clusters, spatially co--located with CO clumps
previously detected by Rubio et~al.\ (2000). Within our analysis uncertainties,
these sub--clusters appear coeval with each other. The most massive stars
appear concentrated in the central sub--clusters, indicating possible mass
segregation. A number of embedded clusters are also observed. This finding,
combined with the overall wealth of dust and gas, could imply that star
formation is still active. An intermediate age star cluster, BS90, formed $\sim
4.3\pm 0.1\, {\rm Gyr}$ ago, is also present in the region. Thus, this region
of the SMC has supported star formation with varying levels of intensity over
much of the cosmic time.
\end{abstract}


\section{Introduction}

The Small Magellanic Cloud (SMC) is an excellent laboratory to investigate the star
formation (SF) processes and the associated chemical evolution in dwarf
galaxies. Its current sub-solar chemical abundance (Z=0.004) makes it the best
local counterpart to the large majority of dwarf irregular (dIrr) and Blue
Compact Dwarf (BCD) galaxies, whose characteristics may be similar to those in
the primordial universe. Because the SMC is part of a triple system, its star
formation can be triggered by interactions with the Milky Way (MW) and the
Large Magellanic Cloud (LMC). This kind of interactions might have been very
frequent in the past, if galaxies are building up via hierarchical mergers.

The SMC proximity allows us to resolve into single stars the youngest and most
compact star clusters, down to the sub solar mass regime. 
Young star clusters have been extensively studied in the MW and in the LMC, and
a direct comparison of the characteristics of these clusters with those of
the SMC allows us to address fundamental issues such as the universality of the
initial mass function (IMF), and the occurrence of primordial mass segregation
within star forming regions.

In this paper we present an in--depth study of the stellar content of NGC~346
($\alpha_{2000}=$ 00:59:05.2; $\delta_{2000}=-$72:10:28). This star cluster is an
outstanding benchmark for star formation studies. It is an extremely young,  
\citep[$\sim$3~Myr,][]{bouret03} moderately compact cluster, that excites
the largest and brightest H{\sc II} region - N66 - in the SMC \citep{relano02}.
N66 has a diameter of about $420\arcsec$, which, at the distance of the SMC
\citep[60.6~Kpc, ][]{hilditch05} corresponds to $\sim 123\, {\rm pc}$.  

NGC~346 lies in a very active and interesting region: at a distance of 
2~arc--minutes to the East (35~pc in projection) of the center of the cluster,
we find the massive Luminous Blue Variable star HD~5980. N66 contains also at
least two known supernova remnants (SNR): SNR~0057-7233 \citep{ye91}, located
to the Southwest of N66, and SNR~0056-7226. SNR~0057-7226 is found in proximity
to HD~5980, although far-ultraviolet absorption studies have determined that
HD~5980 is actually behind the SNR. High--resolution X--ray observations of N66
\citep{naze02} show a $130\arcsec \times 130\arcsec$ region of extended thermal
emission corresponding to the SNR, with an emission peak at its center.
SNR~0057--7226 has a relatively uniform surface brightness and does not show
any clear temperature gradient from center to rim. More details on the
properties of the SNRs and H\,{\sc ii} regions around NGC~346  are given by
\citet{reid06}.

Both H$\alpha$ and O{\sc vi} observations imply the presence of a strong shock
in N66, associated with SNR~0057-7226, as the blast wave encounters the denser
material of the ionized nebula \citep{danfort03}. The dynamical properties of
the region inferred from the emission O{\sc vi} lines observed near NGC~346
suggest the existence of a super--bubble, caused by a supernova explosion
\citep{hoopes02}.

NGC~346 is relatively faint in the X--rays, with a total luminosity of
$L_X\sim1.5\times 10^{34}\, {\rm erg s^{-1}}$ in the 0.3--10.0~KeV energy
range. Most of this emission seems correlated with the location of the
brightest stars in the core of the cluster \citep{naze02}. Many other X-ray
sources are detected in the field of N66. Optical counterparts of these sources
are B--type stars, suggesting that many of them may be X--ray binaries
\citep{naze03}.

NGC~346 contains a major fraction of the O~stars known in the entire SMC
\citep{walborn78, walborn86, niemela86, massey89}. The bright end of its
stellar population ($V\la 19.5\, {\rm mag}$) has been well investigated in the
past twenty years. Evidence of very--early spectral type stars was found 
\citep{niemela86,walborn86,massey05,heap06}, indicating the presence
of stars as young as $\sim 1\, {\rm Myr}$, and an IMF slope ($\Gamma = d\log \xi
(\log m)/d \log m =-1.9$, where $\xi (\log m)$ is the number of stars born per
unit logarithmic mass interval per unit area), close to the slope for massive
stars observed both in the LMC and in the solar neighborhood (Massey, Parker, \&
Garmany  1989, see also Chabrier 2005 for a review). Sequential star formation,
starting from the Southwest of the association towards the center of the cluster
was suggested by \citet{massey89}. 

From the spectral analysis of some dwarf O-type stars, the derived mean
metallicity of the cluster is $Z/Z_{\odot}=0.2 \pm 0.1$ \citep{haser98,
bouret03}, while the estimated age is $t\sim 3 \times 10^6\, {\rm yr}$. In a
more recent work, from the analysis of 21 bright stars, \citet{mokiem06}
identified two different stellar components: though the majority of the stars
are 1--3~Myr old, some of the investigated stars seem somewhat
older (3--5~Myr). No relation between the age and the spatial distribution was found
of the stars.

Extended CO clouds \citep{rubio00} are found within NGC~346. Ground-based
narrow-band images show evidence of dust and compact embedded clusters
surrounding the central cluster, suggesting the presence of a second stellar
generation possibly triggered by the outflows of the central ionizing cluster,
as observed in 30~Doradus and N11 \citep{walborn92,walborn02}.

A rich population of pre--Main Sequence (pre--MS) stars has been recently
discovered by Nota et~al.\ (2006--- hereafter Paper~I) in NGC~346. These stars
have likely formed within NGC~346, $\simeq 3-5\, {\rm Myr}$ ago. They are in
the mass range $3-0.6\, {\rm M_\odot}$, and appear mostly concentrated in the
main cluster and in the knots of molecular gas identified by \citet{rubio00}.

Our ultimate objective is to study how star formation occurred and propagated
in this region, by characterizing its stellar content, luminosity and mass
function. In this paper we present the photometric data, and a qualitative
description of the stellar population in NGC~346. We discuss the properties
and the spatial distribution of the stellar populations identified, and we
propose a preliminary assessment of the star formation history in the region.
The paper is organized as follows: a description of the observation and the
data reduction procedure is presented in Section~2; in Section~3 we present
the color--magnitude diagrams (CMDs), the description of the stellar
populations identified and of their spatial distribution. In Section~4 we
discuss the rich BS90 star cluster in the SMC foreground of NGC~346. The
results of this paper are discussed in Section~5.

\section{Observations and Data Reduction}

\subsection{The data}
\label{data}

Multiple images of NGC~346 were obtained in 2004 July using the Wide Field
Channel (WFC) of the Advanced Camera for Surveys (ACS) on board the HST
(GO-10248, PI: A.~Nota). Several exposures were taken through the filters ${\rm
F555W}$~(V), and ${\rm F814W}$~(I). For a detailed description of ACS filters
the reader should consult the ACS Instrument Handbook \citep{pavlovsky04}.

The observations were designed as follows:  images were obtained at three
different pointing positions (covering an area of $200\arcsec\times 200\arcsec$
each) in the ${\rm F555W}$ and ${\rm F814W}$ filters (see Table~\ref{t:obs}). A
single long exposure was centered on the nominal position of the cluster
(central pointing). The other two pointings were spaced $\sim$1$\arcmin$ from
the center of the cluster toward the North and toward the South, respectively.
For both the northern and southern pointings a dither pattern was especially
designed to allow for hot pixel removal, and to fill the gap between the two
halves of the $4096 \times 4096$ pixel detector. This dithering technique also
improved both point--spread function (PSF) sampling and photometric accuracy by
averaging the flat--field errors, and by smoothing over the spatial variations
in detector response (see Table~\ref{t:obs} and Fig.~1 in Paper~I).

In addition, two short ($2-3\, \rm s$) exposures were obtained for both the
northern and southern pointings, in the two filters, to recover the photometric
information for the brightest sources.

All the images were taken with a gain of $2\, \rm e^- \, \rm{ADU}^{-1}$. The
entire data set was processed adopting the standard STScI ACS calibration
pipeline (CALACS) to subtract bias level, super-bias, and super-dark, and to
apply the flat-field correction. For each filter, all the long exposures were
co--added using the multidrizzle package \citep{koekemoer02}. This process also
provided a correction for geometrical distortion. Cosmic rays (CRs) and hot
pixels were removed. The data units were converted from electrons to
electron--rates, and a final output image was obtained, sampled on a finer grid
(the final pixel scale is $4/5$ of the original pixel dimension, equivalent to
$0\farcs 039$). The pixel scale of the short exposures remained unchanged.
 
The final exposure time of the deep ${\rm F555W}$ and ${\rm F814W}$ images in
the central region is $\simeq 4100\, \rm sec$ and the area is $\simeq 5\arcmin
\times 5\arcmin$, corresponding to $\simeq 88\times 88\, \rm pc$. 

\subsection{Photometric reduction}
\label{photometry}

The photometric reduction of the images has been performed with the DAOPHOT
package within the IRAF\footnote{IRAF is distributed by the National Optical
Astronomy Observatories, which are operated by AURA, Inc., under cooperative
agreement with the National Science Foundation} environment.

We have followed the same methodology used for R136 \citep{sirianni00} and
NGC~330 \citep{sirianni02}, and used PSF fitting and aperture photometry
routines provided within DAOPHOT to derive accurate photometry of all stars in
the field. Given the different crowding properties of the long and the short
exposures, we have performed PSF fitting photometry only on the mosaics
obtained combining the long exposures, and aperture photometry on the images
generated by the short exposures.

\vspace{0.5cm}\noindent$\bullet${\it Short exposure photometry} \\First, we
applied the automatic star detection routine DAOFIND to the ${\rm F555W}$ short
mosaic to identify all the bright sources present in the image. In order to
characterize the appropriate parameters that DAOFIND uses as selection
criteria, we carefully selected by eye a sample of bona fide stars. We ran
DAOFIND with a detection threshold set at $4 \sigma$ above the local background
level, and then we performed a visual inspection of the detected objects to
reject features that had been misinterpreted as stars by the algorithm, such as
spikes from saturated stars.

The stellar flux was measured in an aperture of $0\farcs15$ radius, while the
background was measured in an annulus located between $0\farcs8$ and $0\farcs9$
from the source peak.

The identification of the stars in the ${\rm F814W}$ short image was then
forced assuming the final position of the stars detected in the ${\rm F555W}$
image.

Stars brighter than $m_{\rm F555W}<13$, and $m_{\rm F814W}<13$ are saturated.
With the use of gain setting of 2~e$^-$ ADU$^{-1}$, the WFC CCDs remain linear
well beyond the physical saturation level \citep{gilliland04}. In order to
recover some photometric information for the saturated stars, we performed
aperture photometry on the saturated stars with an aperture of 10~pixels,
following the procedure recommended by \citet{sirianni05}.

The photometric calibration in the Vegamag system was performed by converting
the magnitudes of the individual stars to an aperture radius of $0\farcs5$ and
by applying the zero--points listed in \citet{sirianni05}. These are 25.724,
and 25.501 for the filters ${\rm F555W}$ and ${\rm F814W}$, respectively.

\vspace{0.5cm}\noindent$\bullet${\it Long exposure photometry}\\ In the long
exposures, stars were independently detected in each filter using the DAOFIND
routine, with a detection threshold set at $4 \sigma$ above the local
background level. Their fluxes were measured by aperture photometry with an
aperture of size $0\farcs12$.

Because of the noticeable crowding in the long exposures, we performed
PSF-fitting photometry to refine the photometric measurements of the individual
sources.

In the ACS/WFC, the detector charge diffusion induces variations in core width
and shape of PSF across the field of view \citep{krist03,sirianni05}. In order
to take into account these variations it is necessary to compute a spatially
variable PSF: 160 isolated and moderately bright stars, uniformly distributed
over the entire image, were selected to properly sample the PSF. Stars brighter
than $m_{\rm{F555W}}<18.0$, and $m_{\rm{F814W}}<18.0$ were
discarded from the catalog, because of saturation. As for the short exposures,
fluxes of individual stars were converted to an aperture of $0\farcs5$ and
appropriate zero--points were then applied. All the catalogs were
cross-correlated and merged together. The final catalog contains the photometry
obtained from both the short ($m_{\rm{F555W}}<18.0$ and $m_{\rm{F814W}}<18.4$)
and long ($m_{\rm{F555W}}\ga 18.0$, and $m_{\rm{F814W}}\ga 18.4$) exposures.
The final electronic catalog is shown in Table~\ref{t:cata}.

\subsection{Photometric errors}
\label{error}

\begin{figure}
\epsscale{.9}
\plotone{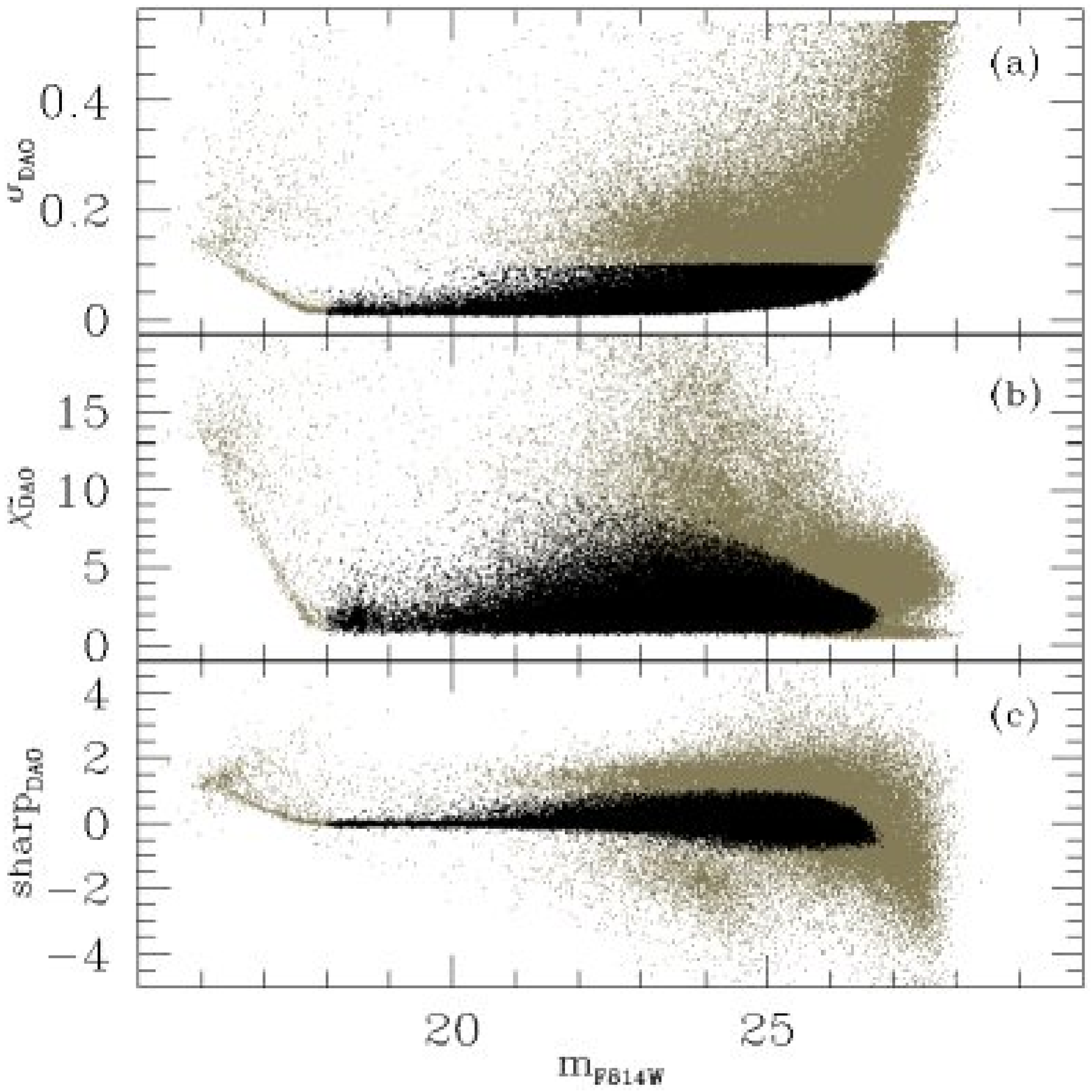}
\caption{\label{f:err} DAOPHOT photometric errors $\sigma_{DAO}$ {\it (a)},
$\chi^2$ {\it (b)} and sharpness {\it (c)} plotted as a function of calibrated
$m_{\rm F814W}$ magnitude (grey dots). The sources with $\sigma_{DAO}<0.1$ are
indicated in the three plots with black dots.} 
\end{figure}

For a safe interpretation of the characteristics of the observed stellar
population, we need to distinguish true single stars from blended and/or
residual spurious objects. Thus, we applied to our catalog selection
criteria based on the quality of the photometric errors. Figure~\ref{f:err}
shows the distributions of the DAOPHOT photometric errors
($\sigma_{\rm{DAO}}$), $\chi^2$ and sharpness as a function of the magnitude
(grey dots) in the ${\rm F814W}$ filter. For each filter, we selected stars
with $\sigma_{\rm {DAO}}<0.1$. We found that, once such a threshold on the
DAOPHOT photometric error had been set, it was not necessary to apply further
selection criteria, based e.g., on shape of the objects (as the \textit{sharpness}), 
or on the ratio between the observed and the expected residuals
obtained with the PSF-fitting (as the $\chi^2$ value). 

In summary, 79,960 sources were found to have $\sigma_{\rm {DAO}}<0.1$ both in
the ${\rm F555W}$ and ${\rm F814W}$ filters.

\subsection{The CTE correction}
\label{cte}

All CCDs on the HST suffer a progressive degradation of their charge transfer
efficiency (CTE) due to cosmic radiation, which causes the partial loss
of signal when charges are transferred down the chip during the readout. The
total amount of charge lost increases with the number of pixel transfers. CTE
degradation can lead to photometric inaccuracy, since the position of a star on
the chip may affect its recorded flux.

A parametric correction for the CTE losses for the WFC is published in the ACS
Data Handbook \citep{pavlovsky04}. This photometric correction depends on the
signal of the object, the background level, the number of charge transfers and
the epoch of the observation.

This parameterization cannot be directly applied to our data, because, due to
the complexity of our mosaic pattern (see Tab.~\ref{t:obs}), the same star
lands in different positions in the different exposures, and, therefore, is
subject to very different CTE effects. We, therefore, developed a special
procedure \citep{sabbi06}. 

First, for each filter, on each single exposure, previously
corrected for geometric distortion, we performed aperture photometry with a
radius of three pixels. The CTE parameterization, as described in the ACS Data
Handbook, was then applied to each photometric catalog, taking into account
the appropriate coordinate transformations, the exposure time, etc. All frames
were aligned to the same coordinate system. Then, for each star in our final
photometric catalog, we computed and applied an average CTE correction. The
typical CTE correction applied to our data is $\sim$0.007 for the F555W
filter, with a maximum value of $\sim$0.013 at $m_{\rm F555W}>26$, and 
$\sim$0.009 for the F814W filter, with a maximum value of $\sim$0.015 at $m_{\rm
F814W}>25$. 

A comparison between our F555W photometry and that obtained in the V~band
by \citet{massey89} shows, on average, a difference of 0.03 magnitude, in
agreement with the quoted errores between the two catalogs. We used Massey's
catalog also to determine the astrometric solution of our own catalog. 

\subsection{The Photometric Completeness}
\label{compl}

Artificial Star experiments are a standard procedure to test the level of
completeness of photometric data. The experiment consists of adding onto a
frame ``artificial'' stars obtained from the scaled PSF used in the photometric
analysis of the frame. Then the artificial stars are retrieved. In total, more
than 2,000,000 artificial stars were simultaneously simulated in the ${\rm
F555W}$ and ${\rm F814W}$ deep exposures, and another 1,0000,000 were added to
the short frames. 

We used the sub--routine ADDSTAR in DAOPHOT to add the artificial stars; the
routine inserts artificial stars of a user--specified magnitude in a copy of
the original frame. We started from the ${\rm F555W}$ image. The artificial
stars were distributed randomly in magnitude with a step function. This
function extends two magnitudes below the detection limit of our observations,
to probe with sufficient statistics the range of magnitude where
incompleteness is expected to be most severe. 

The most intriguing challenge of this experiment is to avoid that stars added
artificially increase the crowding condition of the analyzed frame. To avoid
this potentially serious bias, for both the short and long combined images, we
have divided the ${\rm F555W}$ frame into a grid of sub--images of known
width, and we have randomly positioned only one star in each sub--image. Each
artificial star must have a distance from the sub--image edges large enough to
guarantee that all its flux and background measuring regions fall within the
sub--image, and thus do not contaminate the contiguous artificial star
\citep[see also][]{tosi01}. At each run, the absolute position of the grid
with respect to the frame is randomly changed, guaranteeing that the
artificial stars are uniformly distributed in coordinate space. 

The frame where the artificial stars were added was reduced exactly as the
original frame. We determined the level of completeness of the photometry by
comparing the list of artificial stars added with the one of the stars
recovered. We applied this procedure both to the deep and the short combined
images.

\begin{figure}
\plotone{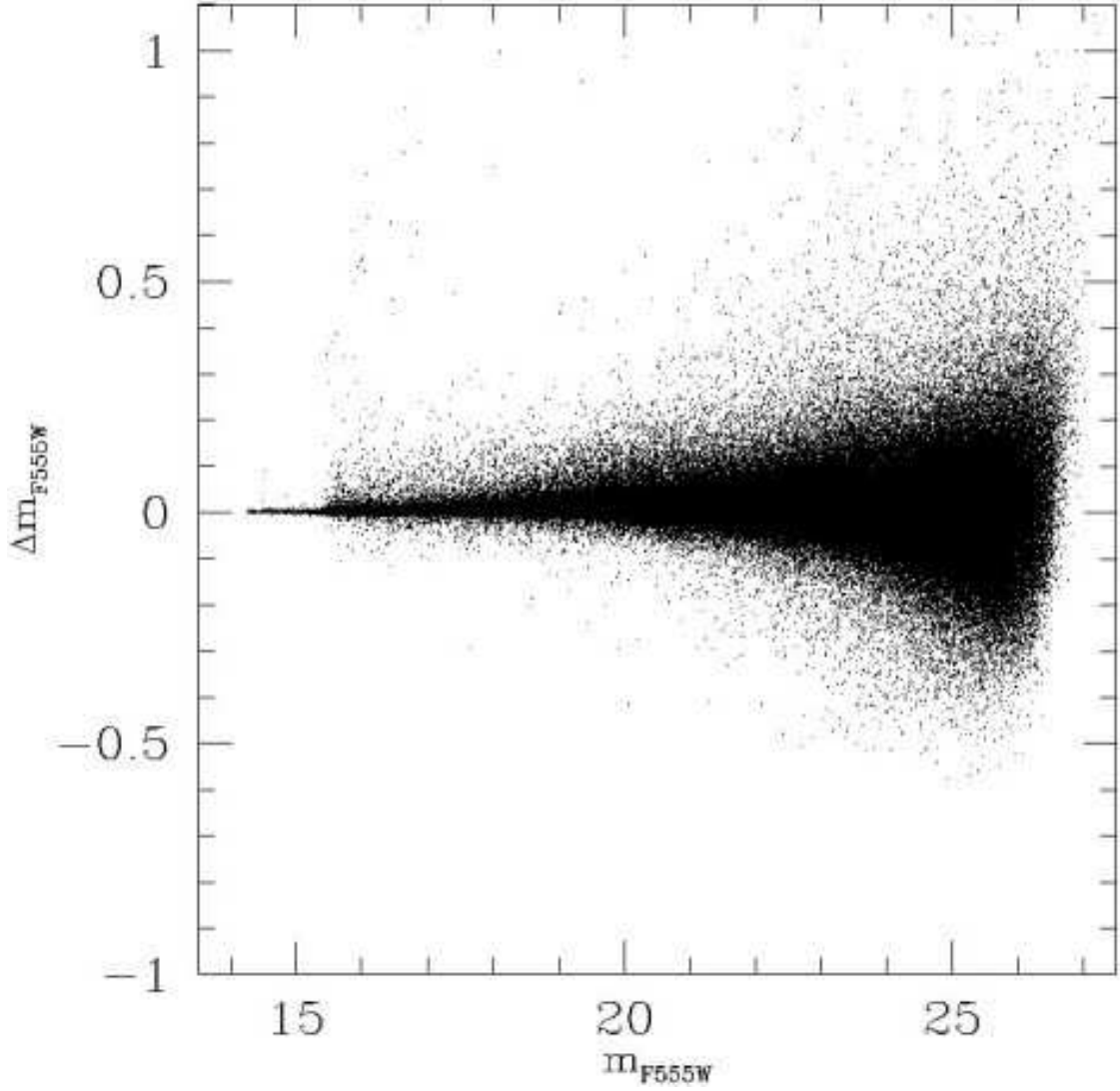}
\caption{\label{f:deltam} Diagram showing the $\Delta$ input-output magnitude
as a function of input magnitude from the Artificial Stars experiments carried
out on the ${\rm F555W}$ image, for artificial stars recovered with
$\sigma_{DAO}<0.1$ in both filters.}
\end{figure}

In our experiment an artificial star is considered recovered if it is found in
both frames with an input-output magnitude difference $\Delta m\le 0.75$ (see
Fig.~\ref{f:deltam}), and at the same time satisfying all the photometric
selection criteria (see \S~\ref{error}). 

\begin{figure}
\plotone{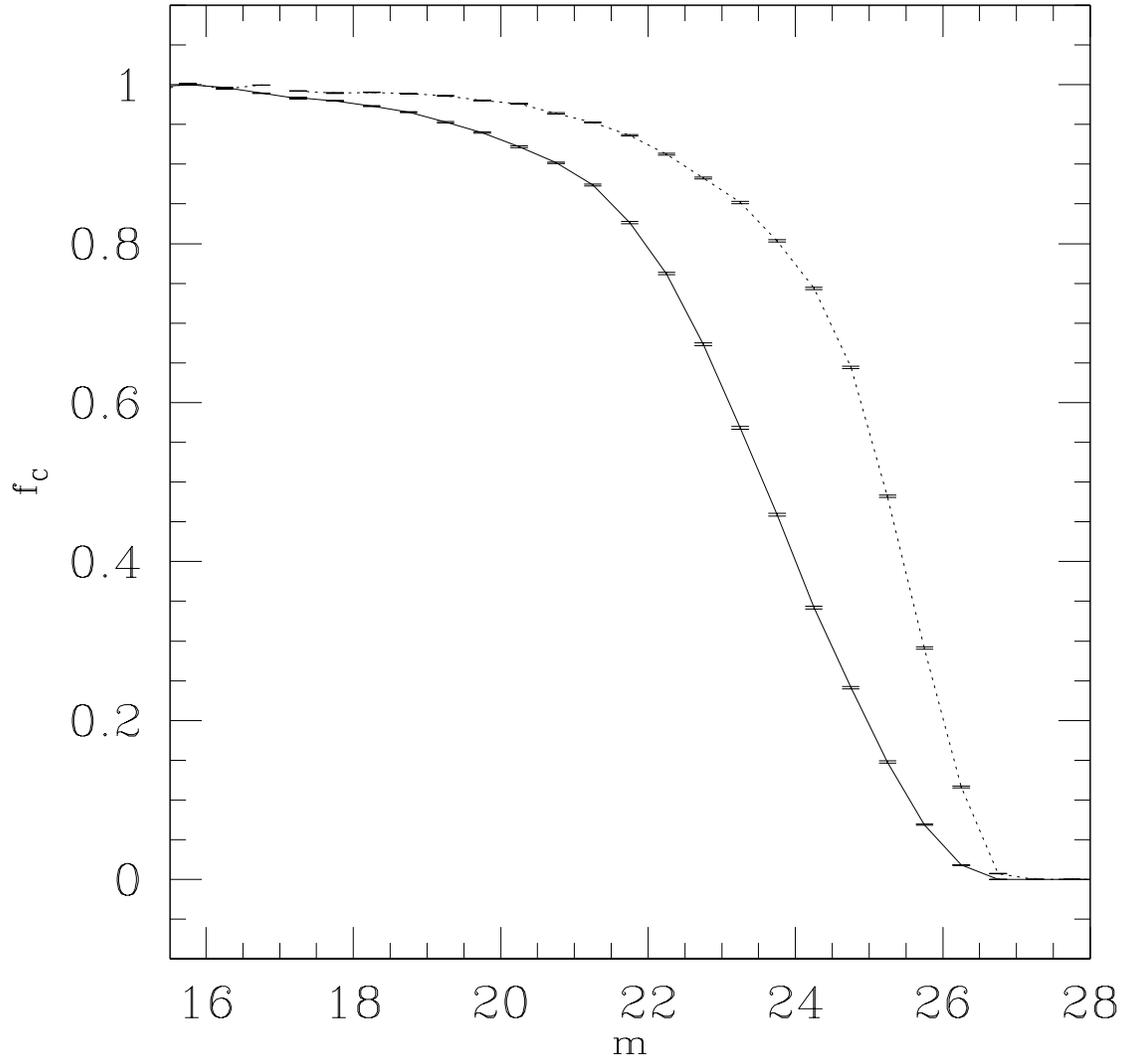}
\caption{\label{f:fc} Completeness curves for the ${\rm F555W}$ (dashed line),
and for the ${\rm F814W}$ (continuous line) photometry as a function of
magnitude.}
\end{figure}

Figure~\ref{f:fc} shows the completeness factor in each filter, defined as the
percentage of the  artificial stars successfully recovered compared with the
total number of stars added to the data: the completeness is better than 50\%
down to $m_{\rm F555W}=25.19$ and $m_{\rm F814W}=23.57$.  

\section{The Color Magnitude Diagrams}
\label{cmds}

\begin{figure}
\plotone{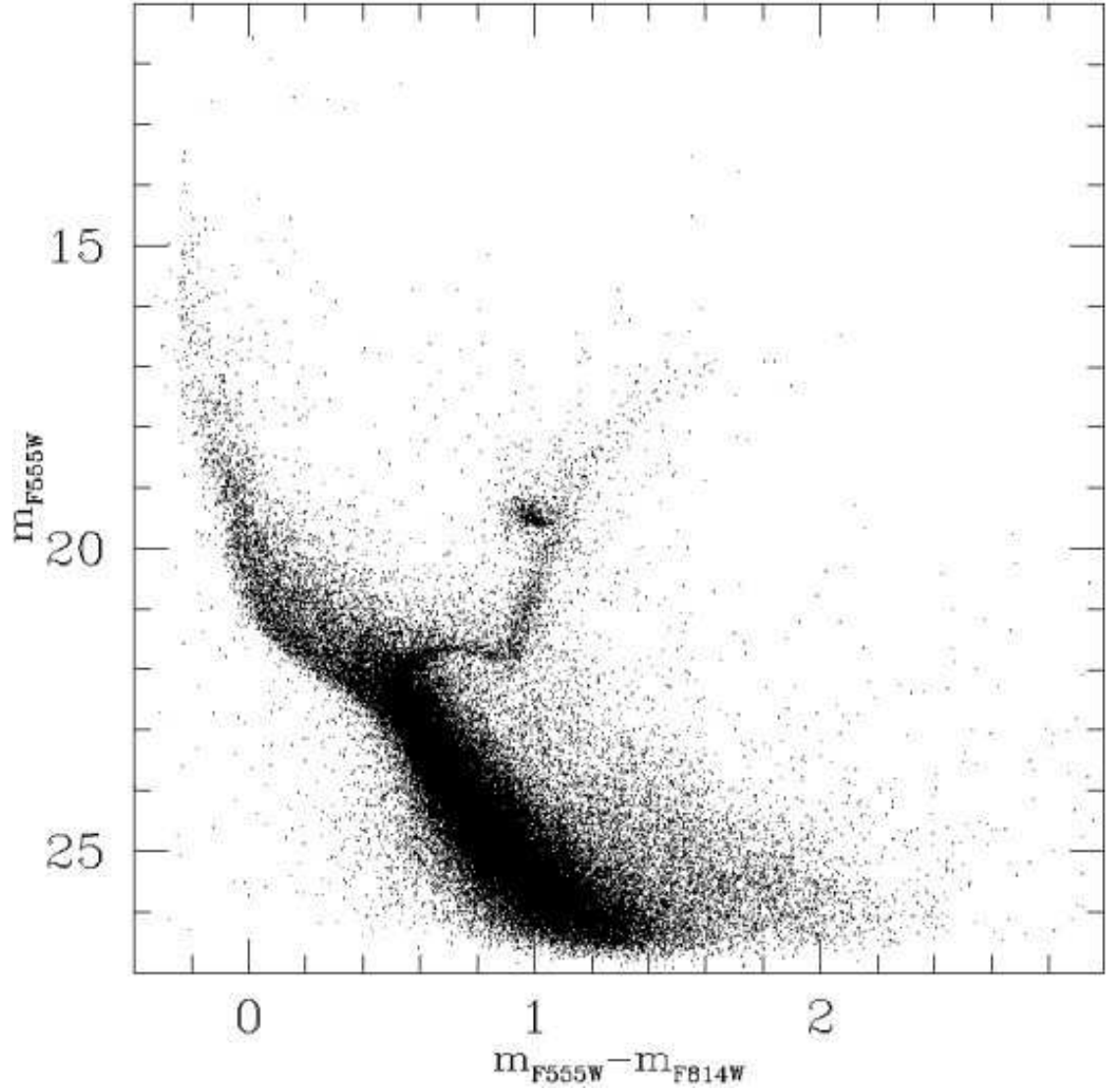}
\caption{\label{f:cmd_all} NGC~346 CMD $m_{\rm F555W}$ vs.\ $m_{\rm
F555W}-m_{\rm F814}$ for all the stars with associated error smaller than 0.1
in both filters.}
\end{figure}

First results on the stellar content of NGC~346 were presented in Paper~I.
Figure~\ref{f:cmd_all} shows the $m_{\rm F555W}$ vs.\ $m_{\rm F555W}-m_{\rm
F814W}$ Color--Magnitude diagram (CMD) of all the stars with
$\sigma_{DAO}<0.1$ mag found in the combined frames. An electronic version of
the photometric catalog is given in Table~\ref{t:cata}, which will be made available, 
and a subset of it is shown in this paper.

A first inspection of this CMD reveals that different stellar populations are
present in the area: 

\vspace{0.5cm}\noindent$\bullet$ {\bf Young stars}: A quite young stellar
population is indicated  by a bright ($12.5\la m_{\rm F555W}\la 22$) and blue
($-0.3\la m_{\rm F555W}-m_{\rm F814W}\la 0.4$) main sequence (MS), well visible
to the upper left of the CMD.

A red ($1.5\la m_{\rm F555W}-m_{\rm F814W}\la 2.2$), faint ($m_{\rm F555W}\la
21$) and well--populated sequence is also clearly visible in the faint and red
part of the CMD. Stars in this region have colors and magnitudes consistent
with being pre-MS stars in the mass range $0.6-3\, {\rm M}_\odot$, which were
formed with the rest of the cluster (Paper~I), $\sim 3-5\, {\rm Myr}$ ago.

\vspace{0.5cm}\noindent$\bullet$ {\bf Intermediate and old age stars}:  An
older stellar population is easily distinguishable in the CMD. Its rich MS
extends from $m_{\rm F555W}\simeq 22$ down to $m_{\rm F555W}\simeq 26.5$. The
evolved phases of this population are well delineated: a narrow sub giant
branch (SGB) is visible at $m_{\rm F555W}\simeq 21.6$ between $0.45\la m_{\rm
F555W}-m_{\rm F814W}\la 0.95$. A red giant branch (RGB) is very well defined
with the brightest stars at $m_{\rm F555W}\simeq 17.3$ and $m_{\rm
F555W}-m_{\rm F814W}\simeq 0.65$. The red clump (RC) of this population is
visible at $m_{\rm F555W}\simeq 19.5$. 

The relative thinness of the SGB suggests that the majority of the older stars
in this field were formed in a single star formation episode that happened
approximately $\sim$4--5~Gyr ago. In \S4 we show that this result is due to the
dominance of the star cluster BS90 over much of our field of view. The
scattering of stars at fainter and brighter magnitudes than the cluster SGB
indicates that star formation has occurred at earlier and later times in this
direction. 

\vspace{0.5cm}\noindent$\bullet$ {\bf Comparison field:} A SMC field ($\alpha
=$ 00:58:42.5; $\delta = -$72:19:46) was observed with WFC for comparison. The
observed field is located $\sim9' 16''$ ($\sim 163\, {\rm pc}$ in projection) from NGC~346,
and its $m_{\rm F555W}$ vs. $m_{\rm F555W}-m_{\rm F814W}$ CMD is shown in
Figure~\ref{f:field}. Only stars which in both filters have $\sigma_{DAO}<0.1$
are plotted. In this diagram we can clearly distinguish an old stellar
population, characterized by broad SGB and RGB. Since the comparison field
does not present any influence of the BS90 star cluster, we see clearer
evidence for a wider range in ages and/or chemical abundances among the stars,
possibly also mixed with more significant stellar population depth along the
line of sight. Younger stars in the core helium burning blue loop phase are
visible above the RC ($17\la m_{\rm F555W}\la 19$, $0.5\la m_{\rm
F555W} - m_{\rm F814W}\la 1$) suggesting that star formation was recently
active in this region. 

Although a quantitative characterization of the field population will be
possible only after computing synthetic CMDs, from the CMD morphology observed
we can already infer that in this area a major episode of star formation
occurred between 3~and 5~Gyr ago, but stars with ages up to at least 10~Gyr are
also present. The MS of the field extends up to $m_{\rm F555W} \simeq 15$, but
its brighter part is much less populated and less straight than that of
NGC~346, indicating that at most recent epochs the SF activity in the field has
been significantly lower, with a possible moderate enhancement $\sim 150\, {\rm
Myr}$ ago.

\subsection{Spatial distribution and age of the NGC~346 region stellar
populations}
\label{starpop}

In order to investigate the spatial distribution of the identified stellar
populations, we have divided the image in 16 regions of $\simeq 1\farcm 1
\times 1\farcm 2$ in size, corresponding to $\simeq 19.4\times 21.2$~pc (see
Fig~\ref{f:mappa}). An inspection of the 16~CMDs covering nearly the entire
region (Fig.~\ref{f:cmds}) reveals that the spatial distribution of the various
stellar populations is not uniform. Contamination from the SMC field also is
observable at varying levels in all of the CMDs. We will analyze the Hess
diagrams corrected for the field star populations in a later paper.

The diffuse nebulosity visible in Figure~1 in Paper~I is due to the ionized gas
contribution in H$\beta$ and [O{\sc iii}]. This is visible especially in
regions~10 (which is at the center of the cluster), 11, and 13 (Southeast
dusty region). In region~6, to the North of NGC~346, we clearly detect an
increase of bright and red stars associated with the surprisingly rich BS90
star cluster (see \S\ref{oldGC}) that was cataloged for the first time by
\citet{bica95}.

The brightest ($m_{\rm F555W}\simeq 15.5$) and bluest ($m_{\rm F555W}-m_{\rm
F814W}\simeq -0.2$) stars are concentrated in the CMD of region \#10 (CMD--10)
that corresponds to the center of NGC~346. This region is characterized by the
presence of many bright and compact stellar sub--clusters, which will be
further discussed in \S~\ref{sc}. The brightest stars in this region have colors
and magnitudes consistent with a stellar population of $3\pm 1$~Myr, and a mass
of $\sim 50-60\, {\rm M}_\odot$. As in Paper~I, this CMD shows also the largest
number of candidate pre--MS stars ($\sim$2100 per unit area).

Candidate pre-MS stars are also visible, but in lower numbers, in the CMDs
obtained for the regions around NGC~346 (regions~5, 6, 7, 9, 11, 13, 14, and
15). In particular CMD~11 (which corresponds to the region where we can see the
dusty ``ridge'') and CMD~13 \citep[which corresponds to the region where the
majority of embedded clusters are found, see \S\ref{sc} and ][]{rubio00} show a
young MS nearly as bright ($m_{\rm F555W}\simeq 14-15$) as that in CMD~10,
indicating that the star formation has been, and possibly still is, active in
this region.  

Region~6 is centered on the core of the star cluster BS90. The presence of this
intermediate age stellar population is easily distinguishable in the
corresponding CMD (CMD~6 in Fig.~\ref{f:cmds}): compared with all the other
CMDs, the old MS appears more populated ($\ga 12100$ stars per unit area) and
both the RGB and SGB are better delineated and narrower. The presence of this
old stellar association is visible also in the surrounding regions (3, 5, 10,
11, and probably, 2, 7, 9).

CMD~4 corresponds to an external region located toward the East, at a distance
of $\sim 141\arcsec$ ($\simeq 41.4$~pc) from the center. In this area the blue
MS appears brighter ($m_{\rm F555W}\simeq 12.5$ in projection) than those of
regions~1, 2 and~5, but pre--MS stars as young as those identified in the
innermost CMDs are not present. The brightest and bluest stars detected in this
area belong to a small stellar association, which is probably older, and likely
not connected with NGC~346 (see \S~\ref{sc}). 

\begin{figure}
\plotone{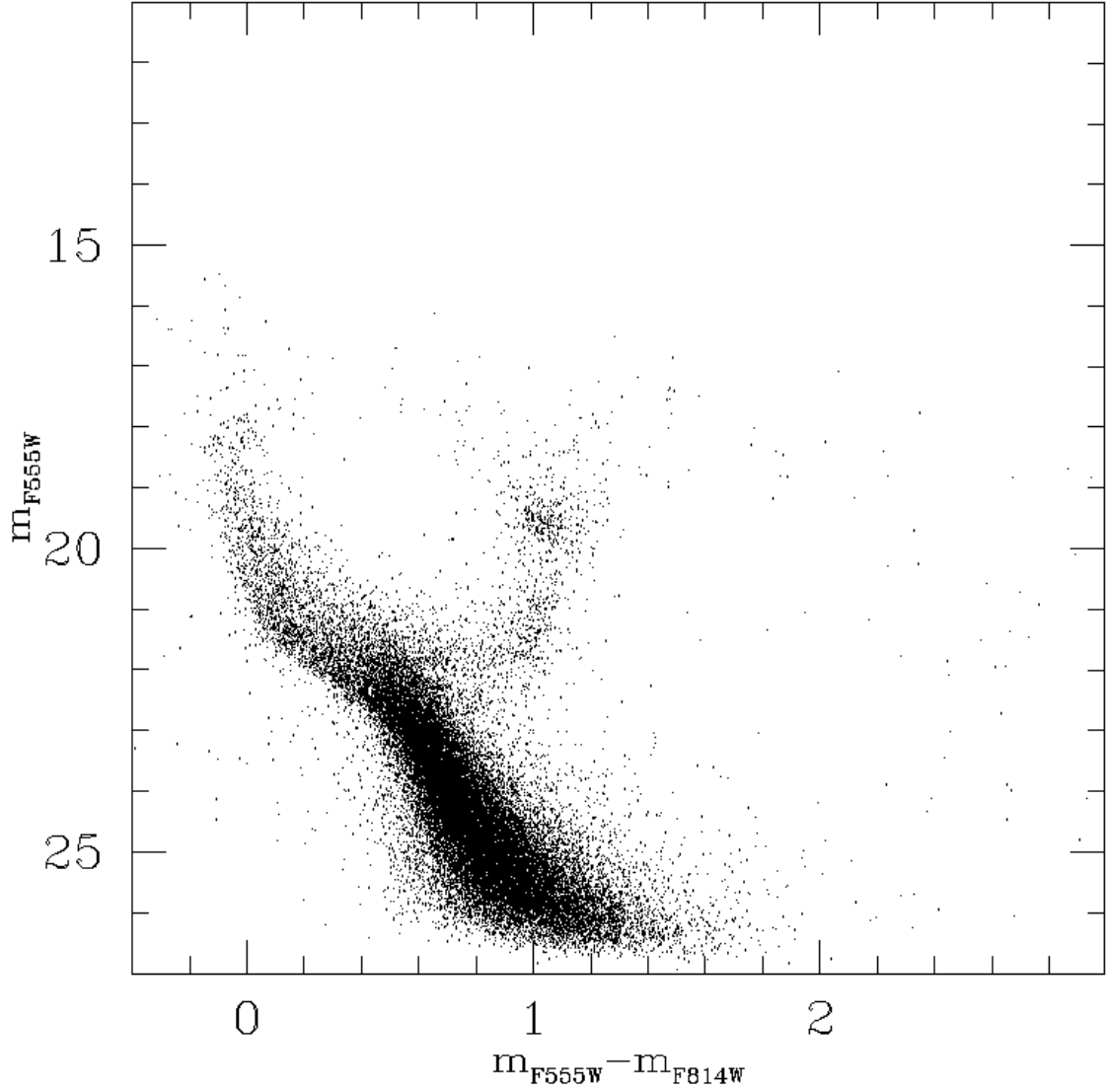}
\caption{\label{f:field} CMD of a SMC field region observed for comparison. The
same error selection criteria used for NGC~346 were applied to this photometric
catalog.}
\end{figure}

\begin{figure}
\plotone{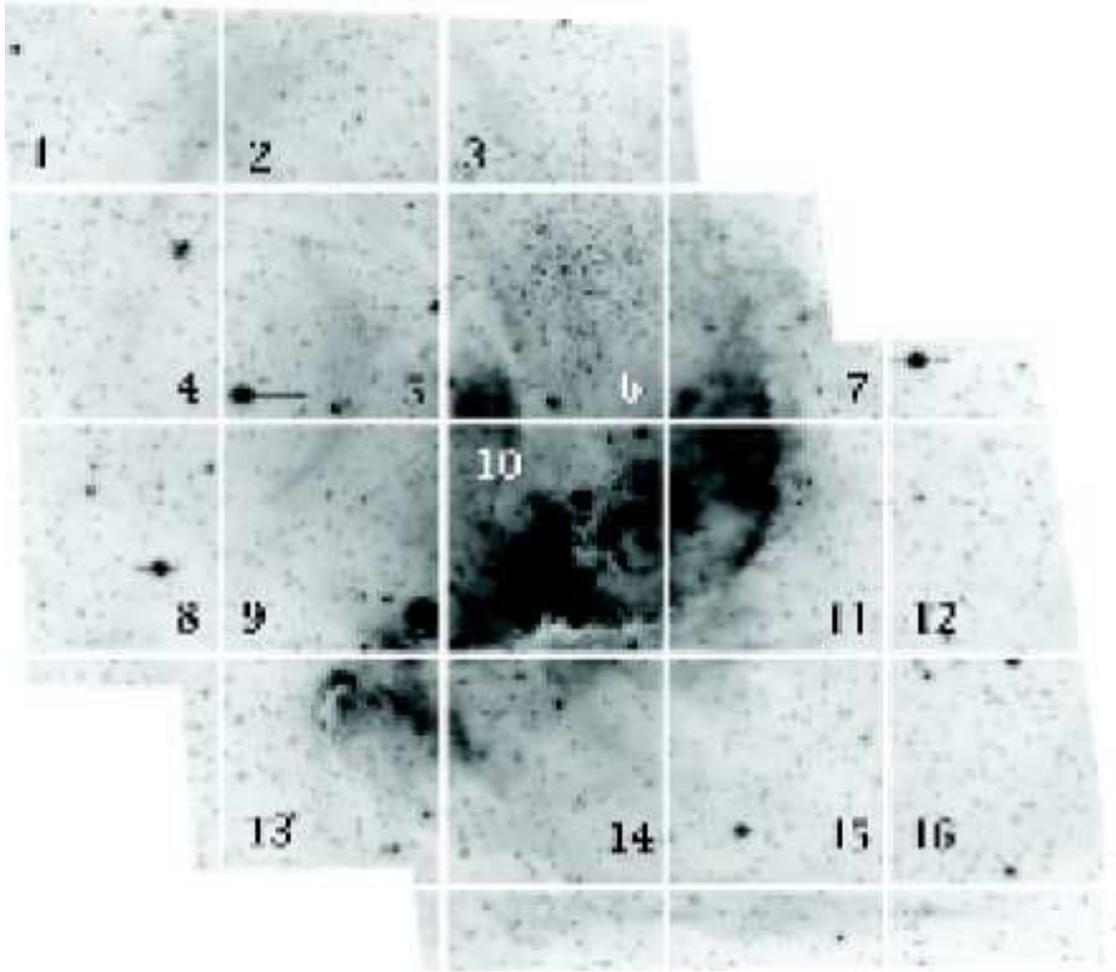}
\caption{\label{f:mappa} NGC~346 in the light of $\rm{F555W}$. The image was
divided into 16 regions of $\simeq 0\farcm3$ in size to study the spatial
distribution of the various stellar populations.}
\end{figure}

\begin{figure}
\plotone{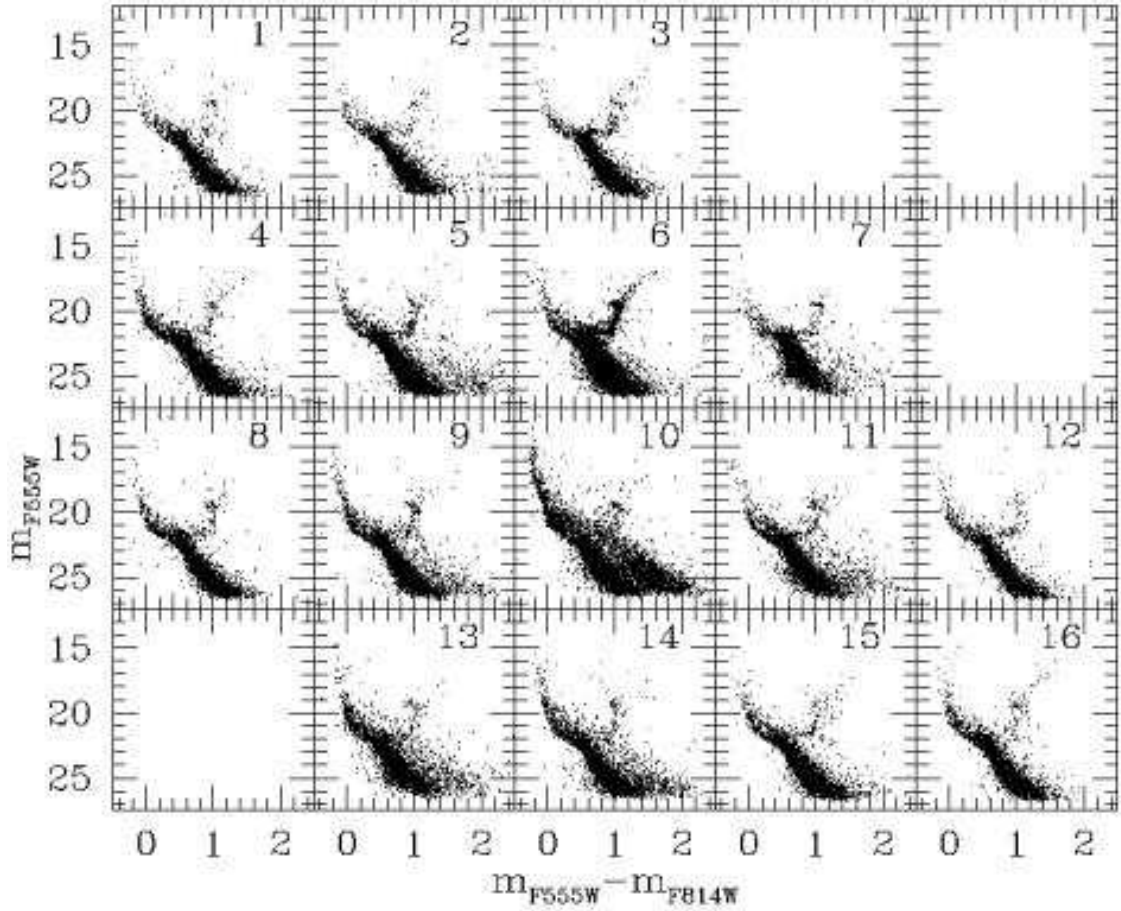}
\caption{\label{f:cmds} CMDs of the stars measured in the ACS FoV and selected
on the basis of their position, indicated in Figure~\ref{f:mappa}.}
\end{figure}

CMDs~1, 2, 8, 12, and 16, are at the outskirts of NGC~346, and show only the
SMC field population (e.g. to be compared with the CMD in Fig.~\ref{f:field}):
in all these CMDs both the RGB and the SGB appear broader, and less well
delineated than those observed in the innermost CMDs. Also the RC is less
clearly defined. The blue MS in these CMDs ends at $m_{\rm F555W}\simeq 15$, as
observed in the case of the comparison field.  

We used the set of Padua isochrones \citep{fagotto94}, transformed into the
VEGAMAG system by applying the transformations calculated by
\citet{origlia00}, to derive a preliminary evaluation of the ages of the
stellar populations identified. For this calculation we assumed a distance
modulus $(m-M)_0=18.9$ and a Galactic foreground $E(B-V)=0.08$.

The blue MSs observed in the CMDs are likely to be the composition of multiple
star formation episodes. In all the CMDs we found a young stellar population,
belonging to the SMC field (see also Fig.~\ref{f:field}). Superimposed on this
population we found a $\sim 15\pm 2.5\, {\rm Myr}$ old stellar population in
CMD~4, and a very young stellar population (with an age $\sim 3 \pm 1$~Myr) in
CMDs~5, 6, 7, 9, 10, 11, 13, 14, and 15. Similar age estimates are derived for
the pre-MS candidates found in these CMDs. 

\subsection{The young stellar population}
\label{sc}

\begin{figure}
\plotone{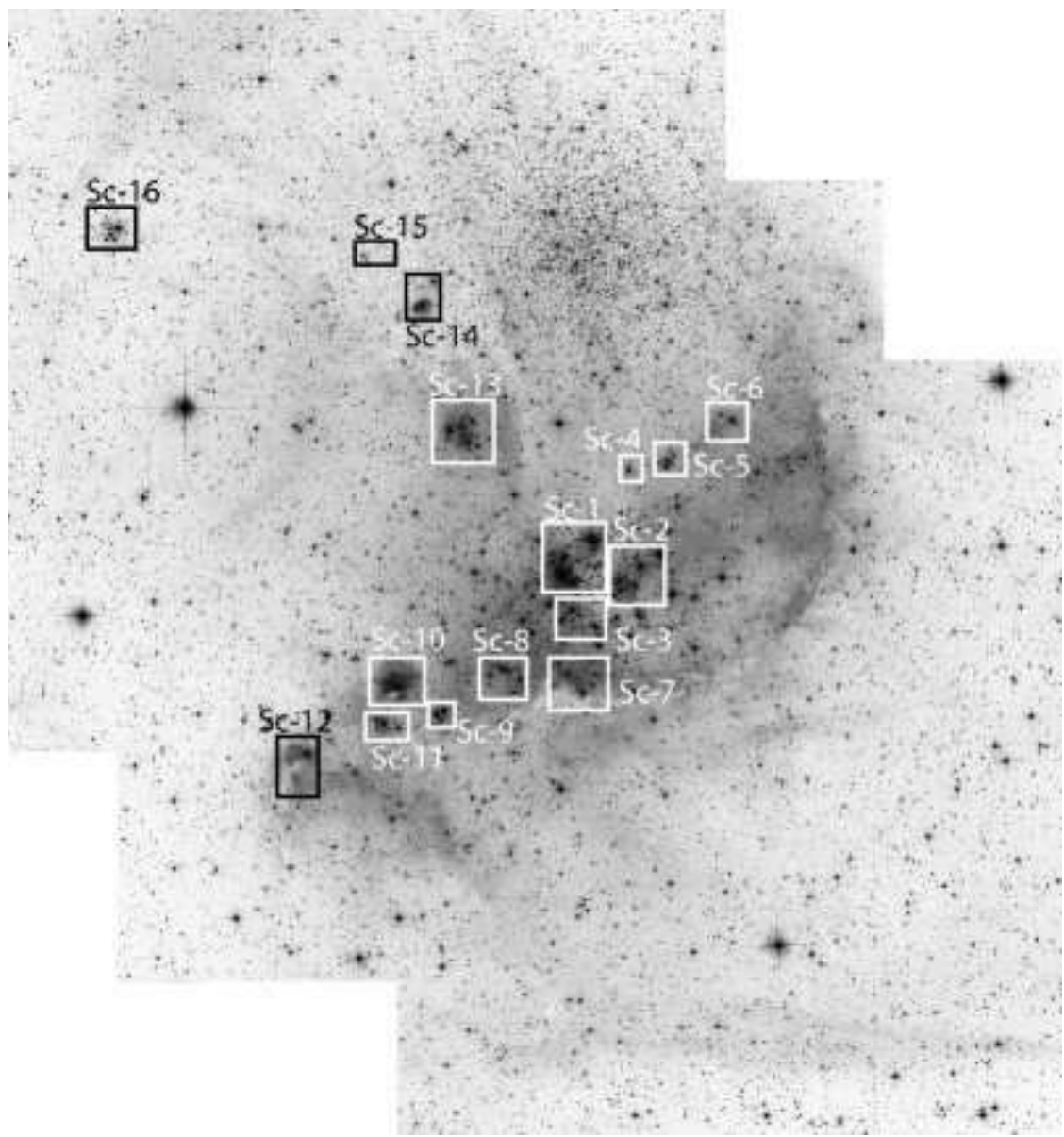}
\caption{\label{f:starc1} Stellar associations identified within the N66
nebula.}
\end{figure}

The youngest stellar population appears concentrated in a number of small
compact stellar associations (see Fig.~\ref{f:starc1}), which vary in density,
dust content and morphology, and are co--located with the dense CO clumps found
in the region by \citet{rubio00}. The sub--clusters were first identified using
isophotal contours; then we considered sub--clusters those stellar associations
which show a stellar density at least three times higher than the average stellar
density.

The majority of these associations are embedded in H {\sc II} gas, whose
distribution is traced (Fig.~\ref{f:starc1}) by the diffuse emission in
H$\beta$ and [O{\sc iii}] 4959, 5007.

In the deep HST/ACS images the central cluster seen in the ground based images
is resolved in three separate sub--clusters (namely Sc--1, 2, and 3 in
Fig.~\ref{f:starc1}). Arcs of dust and gas depart from these three
sub--clusters towards the Northwest and connect them with three smaller
associations (Sc--4, 5, and 6). 

Another 6 sub--clusters (namely Sc--7, 8, 9, 10, 11, and 12) are found toward the
Southeast. Some of these associations (i.e.\ Sc--10, and 12) appear still embedded in
dust and fuzzy nebulosities, and probably are sites of recent or even still
ongoing star formation.

A dense clump of molecular gas is located along the perpendicular direction
from the main body of NGC~346 to the Northeast \citep[see][]{rubio00}. At its
base there is another bright and dense sub--cluster (Sc--13). Another two
faint and red associations (Sc--14, and 15 respectively), constituted almost
exclusively by pre--MS stars, are located at the center of this clump.

Another bright sub--cluster (Sc--16) is located at the extreme Northeast
periphery of the region. This association shows redder colors and does not
display any nebulosity, suggesting an older age, with respect to the other
younger star clusters.

The average characteristics of the sub--clusters are summarized in
Table~\ref{t:sc}. 

How did the star formation progress within the N66 nebula? Are all the stellar
sub--clusters identified here coeval, or, as suggested by \citet{massey89} did
they form following a sequential process? Has N66 exhausted its fuel, or is its
star formation still ongoing?
If so where, and at what levels?

\begin{figure}
\plotone{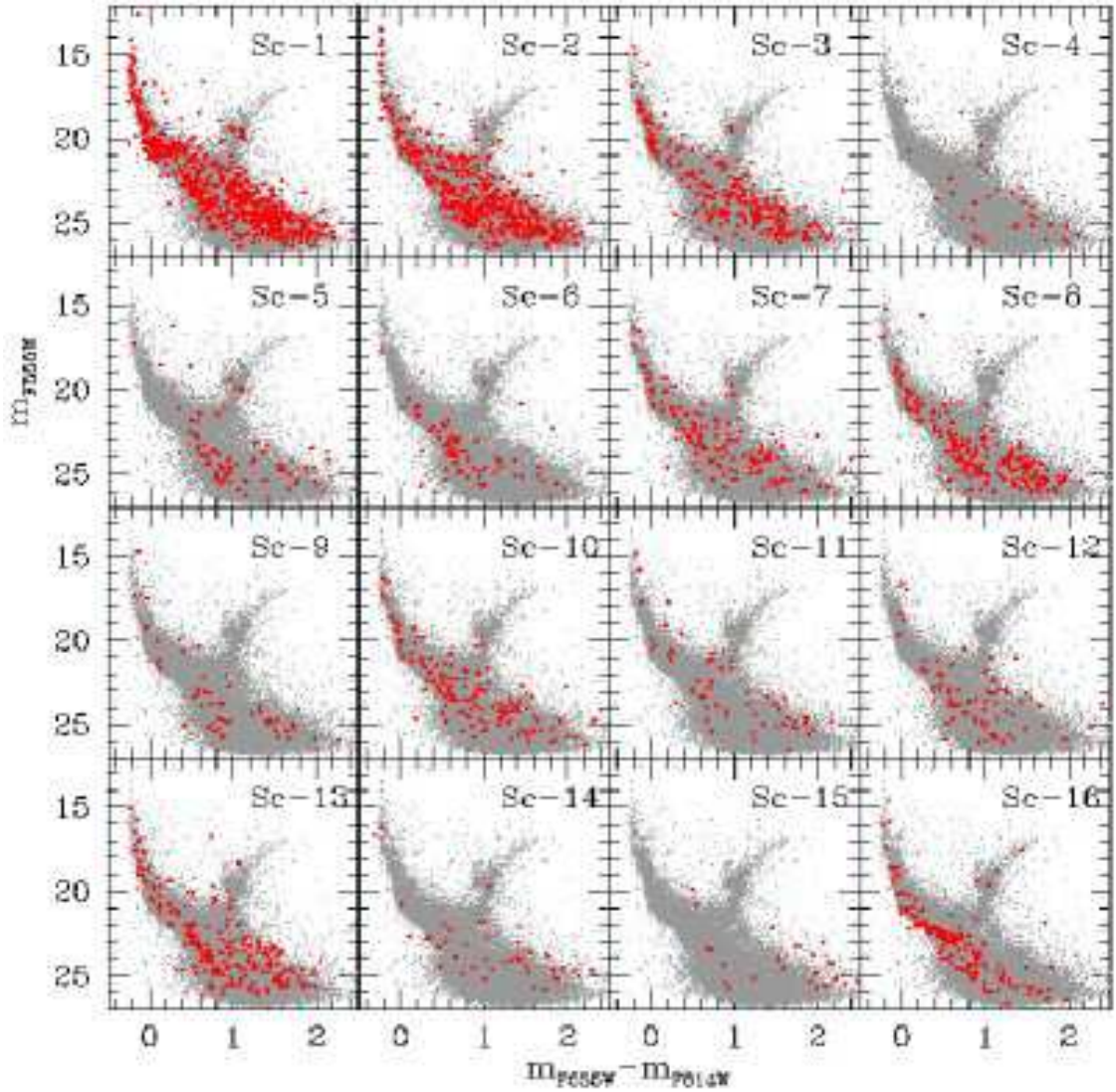}
\caption{\label{f:starc2} CMDs of the 16 sub--clusters identified within
NGC~346 region. Red dots indicate stars within the sub--cluster radius (see
Tab.~\ref{t:sc}). In order to better distinguish the evolutionary phases,
sub--clusters CMDs are superimposed to the total CMD (grey dots).}
\end{figure}

The CMDs of all the young sub--clusters are shown in Figure~\ref{f:starc2}. To
estimate sub--clusters ages and determine how the star formation has progressed
in the region, we fitted the CMDs of those sub--clusters which show a populated
MS (Sc--1, 2, 3, 7, 8, 9, 10, 11, 13, and 16) with Padua isochrones
\citep{fagotto94}. Because the majority of the sub--clusters populations are
constituted by pre--MS stars, we used also pre-MS isochrones by \citet{siess00},
to evaluate the age of the various sub--clusters.

We found that:

\begin{itemize}
\item From the comparison with the Padua isochrones, we estimated an age of
$3\pm 1\, {\rm Myr}$ for the MS of sub--clusters Sc--1, 2, 3 and 13. These 
4~sub--clusters show also a high number of pre--MS stars (see Table~\ref{t:sc}).
A comparison with Pre--MS isochrones \citep{siess00}, indicates ages of $4\pm
1\, {\rm Myr}$ for these stars. This is in good agreement with the age derived
from the MS, taking into account that we are trying to estimate the age of the
sub--clusters using different sets of isochrones, computed for different
evolutionary phases.
\item Compared to Sc--1, 2, 3, and 13, the MS of sub--clusters Sc--7, 8, 9, 10,
and 11 appears redder, suggesting either that these sub--clusters are few
million years older, or that they are coeval with the others, but affected by
higher extinction. However, magnitudes and colors of the pre--MS stars in
sub--clusters Sc--7, 8, and 10 appear too bright and red to be compatible with
a stellar population older than $\sim 4\, {\rm Myr}$, indicating a more likely
age of $3 \pm 1\, {\rm Myr}$ also for these sub--clusters. This last hypothesis
is also supported by the presence of dust and fuzzy nebulosities.
\item Both MS and pre--MS isochrones indicate that the sub-cluster
Sc--16 has an age of $15\pm 2.5\,{\rm Myr}$. It probably is a Pleiades-size
open star cluster that may have formed independently of the main NGC~346 system.
\item Figure~\ref{f:starc2} shows that sub--clusters Sc--5, 6, 12, 14, and 15
are not sufficiently populated to allow a statistically meaningful evaluation
of their ages via MS fitting, but colors and magnitude of the pre--MS stars 
detected in these sub--clusters are in agreement with an age of $4 \pm 1\, {\rm
Myr}$.
\end{itemize}

The crossing time $T_c$ from the central sub--cluster to the external
sub--clusters (namely Sc--6, 12, 15 and 16) for a sound speed of 10 Km/s is of
order of $\sim 2\, {\rm Myr}$, with the exception of Sc--16  where it is
$\simeq 4\, {\rm Myr}$.  

Comparison with isochrones does not allow us to resolve differences in age
smaller than 1--2 Myr. Within this margin, all sub--clusters appear coeval 
with each other. The one exception is Sc--16, that may not be related to the
star-forming episode that originated NGC~346. The most massive and brightest
stars are found in the three innermost clusters (namely Sc--1, 2, and 3).
Moving from the center to the South the star clusters become progressively less
massive. Star clusters towards the N, Northeast, are almost entirely
constituted by low mass pre--MS stars, with only few stars, if any, on the MS.
These sub--clusters are co--located with of the highest velocity CO clouds
observed by \citet{rubio00}; in particular Sc--14 and 15 are located in the
center of the dense CO cloud in the Northeast ``spur''. 

The fact that the North, Northeast star clusters are co--located with the
highest velocity CO clumps could indicate that the stellar winds have started
to disperse the ionized gas on the North side of the nebula (Sc--4, 5, and 6).
Towards the East and the South the gas density is likely high enough to resist
the wind pressure. A multitude of dark Bok--globules, similar in structure to
those in M~16 can be found close to the Southeast clusters, along a dust
``ridge'' which extends toward the South--Southwest. This suggests that N66 has
not yet exhausted all its fuel, and that the star formation process is still
active at some level in the periphery of the central cluster.

\subsection{Star formation timing}

The sequence of star formation in N66 presents an intriguing picture. Only
Sc-16 appears to be old enough to be associated with the event that produced
the larger SNRs in this region of the SMC (see \S1).  Ages of all of the other
stellar groupings fall in the range of 2--5~Myr, implying an age spread, if
present, of $\leq$3~Myr over an area that is about 40~pc in extent.  The N66
region thus lies on the age spread--linear size relationship described by
\citet{elmegreen00}.

\citet{elmegreen00} interprets the age range-size correlation as a result of
the  formation of self-gravitating sub--clumps within 1--2 crossing times of a
larger  cloud complex. The observed sub--clusters in N66 then could be the
products of the unstable sub--clumps required by this model, some of which may
still be collapsing and are seen as the star forming CO-clouds. Following the
discussion of \citet{rubio00}, the diffuse molecular gas in the original cloud
likely has been photodissociated and probably largely photoionized by O~stars
in the main NGC~346 cluster. The remaining molecular sub--clumps are
experiencing rapid evolution due to photoevaporation from their surface PDRs
and varying levels of star formation within the clouds.

In this conceptual model of the N66 region, most of the star formation was not
triggered, but rather resulted from the turbulence driven density variations
within a giant interstellar cloud complex.  It also helps us understand how N66
can have a relatively normal structure which resembles Galactic H{\sc ii}
regions even when the mass loss rates from its O~stars are substantially
reduced relative  to those of its Galactic counterparts \citep{bouret03}.
Mechanical power from O-star winds is not a dominant factor in the evolution
within N66. On the other hand the presence of multiple SNRs surrounding N66
\citep{reid06} suggests that its outer could well be influenced by mechanical
processes.  We will return to a more detailed examination of relationships
between stars and gas in a later paper. 

\section{BS90: A Rich Intermediate Age Star Cluster}
\label{oldGC}

As mentioned in \S~\ref{starpop} an old stellar cluster is present in the
region. It is clearly visible to the North of NGC~346 (see Fig.~1 in Paper~I).
Its CMD is displayed in Figure~\ref{f:rgbs}b. The various evolutionary
features (MS, SGB, RGB and RC) are quite tight, showing that we are
actually dealing with a simple stellar population. The quality of the
photometry is such that we can clearly identify a horizontal gap at $m_{\rm
F555W}\simeq 22.3$, just below the MS Turn--Off (TO). This gap corresponds to
the overall contraction of stars when their central hydrogen fuel is consumed
down to a few percent, and has been observed in Galactic open clusters of a few
billion years \citep[][and references there in]{bragaglia06}. The comparison of
the cluster CMD with the Padua isochrones indicates that the system was formed
around 4~Gyr ago. As noted in \S\ref{cmds} the evolutionary sequences are wider
in the field CMD (Fig.~\ref{f:rgbs}a) indicating that the field also
developed a major episode of star formation between 3 and 5~Gyr ago, but it
probably lasted longer. As a result, stars of different ages and metallicity are
present. 

We have seen in \S\ref{sc} that there is evidence of variable reddening in our
data. Nevertheless both the RC and the SGB of the old star cluster do
not seem to be affected by differential extinction, suggesting that the old
cluster is likely placed in front of NGC~346.

In order to study the characteristics of this system we computed its density
profile. First, we determined the center of gravity ($C_\mathrm{grav}$) of the stellar
association, by estimating the position of the
geometric center of the star distribution within an area of $\sim 4\arcmin 
\times 4\arcmin$ around the center of the cluster. To determine this position
we applied the same procedure described in Montegriffo et~al.\ (1995), which
computes $C_\mathrm{grav}$ by simply averaging the $\alpha$ and $\delta$ coordinates
of stars located in the selected area. The position we obtained for the stellar
association is $\alpha = 00^h59^m04\farcs 59,\, \delta = -72\degr 09^m 11\farcs
73\, [J2000]$, with a 1$\sigma$ uncertainty in both $\alpha$ and in $\delta$ of
$\sim 0\farcs 5$, that corresponds to about 12 pixels in the HST/FWC images.

We computed the star density profile applying the standard procedure described
in \citet{calzetti93}: the entire photometric sample was divided into 15
concentric annuli, centered on $C_{grav}$, spanning a spatial range from
$0\arcsec$ to $158\farcs 5$. Then, we determined the number of stars in each
annulus. For each annulus the star density was obtained by dividing the number
of stars by its area, expressed in arcsec$^2$.

Figure~\ref{f:king} shows the stellar density profile. The open triangles
represent the density profile we obtain if we include all the stars present in
our photometric catalog. The contamination due to the presence of NGC~346 is
clearly visible. In order to reduce the sources of uncertainties we then
re--computed the density profile considering only the stars redder than
V$-$I$\ga 0.4$. Stars bluer than this value belong to a young population, that
can not be coeval, and thus related to the globular cluster. The candidate
pre--MS stars were also removed from the catalog. The final density profile is
also shown in Figure~\ref{f:king} (black dots). The error--bars were computed
using the formula 
$$
\sigma= \sqrt{\frac{n_{\star,j}(1-n_{\star,j})}{A_j}}
$$
\noindent were $n_{\star,j}$ is the number of stars within the j$^{th}$
annulus, where $A_j$ is the area of the j$^{th}$ annulus. 

The solid line represents the best fit of the stellar density distribution of
the old cluster (black dots), and was obtained by applying a King model (1966)
with a core radius $r_c=25\arcsec$ and a tidal radius $r_t=130\arcsec$. The
concentration of the cluster ($c=\log(r_c/r_t)=0.72$) indicates that the cluster
is still far from its gravitational collapse (see Meylan \& Heggie, 1997), in
agreement with its relatively young age ($\sim 4\, {\rm Gyr}$). 

To better estimate the age of the cluster, we applied the synthetic CMD method by
\citet{tosi91} in the updated version described by \citet{angeretti05}. We find
that the synthetic model in better agreement with the data has Z=0.003,
E(B-V)=0.08, distance modulus =18.9, age$=4.3\pm 0.1\, {\rm Gyr}$ and a fraction
of binaries between 30 and 50\%. The comparison between the observed and the
synthetic CMDs of the core of the cluster is shown in Figure~\ref{f:syn}.
Corresponding luminosity functions (LFs) are shown in Figure~\ref{f:LFs}. 

The circumstance that both the old cluster and the field contain a population
$\simeq 4.3\,  {\rm Gyr}$ old shows that in our observed region the star
formation was quite active at that epoch. This result is not consistent with
the simple two burst models proposed by \citet{pagel98} and \citet{rich00}, but
agrees with \citet{gardiner92}, \citet{dolphin01}, and  \citet{mccumber05}, who
suggested that both the SMC field and the old star clusters have probably
formed in a quasi--continuous mode with a broad peak between 4 and 12~Gyr.

The synthetic CMD methods allowed us also to estimate the entire mass content
between $0.6\, {\rm M_\odot}$ and $120\, {\rm M}_\odot$ of the cluster within its
core radius ($\sim$$5.6\times 10^4\, {\rm M}_\odot$). From the King profile, we
estimated that this correspond to $\sim$77\% of the total mass of the cluster.
To take into account the contribution of the sources in the mass range between
$0.05-0.6\, {\rm M}_\odot$, in this mass range, we have assumed an initial mass
function with $\alpha=1.3$ \citep{kroupa03}. In doing this we estimate for the
cluster a total mass of $\sim$$1.03\times 10^5\, {\rm M}_\odot$.
 
\section{Conclusions}

Our analysis of the stellar content of N66/NGC~346 provides a snapshot of the
star formation history of the entire region. From the analysis of the
photometric data, we can conclude that different star formation episodes took
place. 

The oldest stellar population observed belongs to the SMC field. The morphology
of the CMD suggests that, in the field, a major episode of SF occurred
approximately between 3 and 5~Gyr ago, but stars with ages up to at least 10~Gyr 
are also present. We noted also that at most recent epochs the SF activity
in the field has been significantly lower, with a possible moderate enhancement
$\sim$150~Myr ago.

\begin{figure}
\plotone{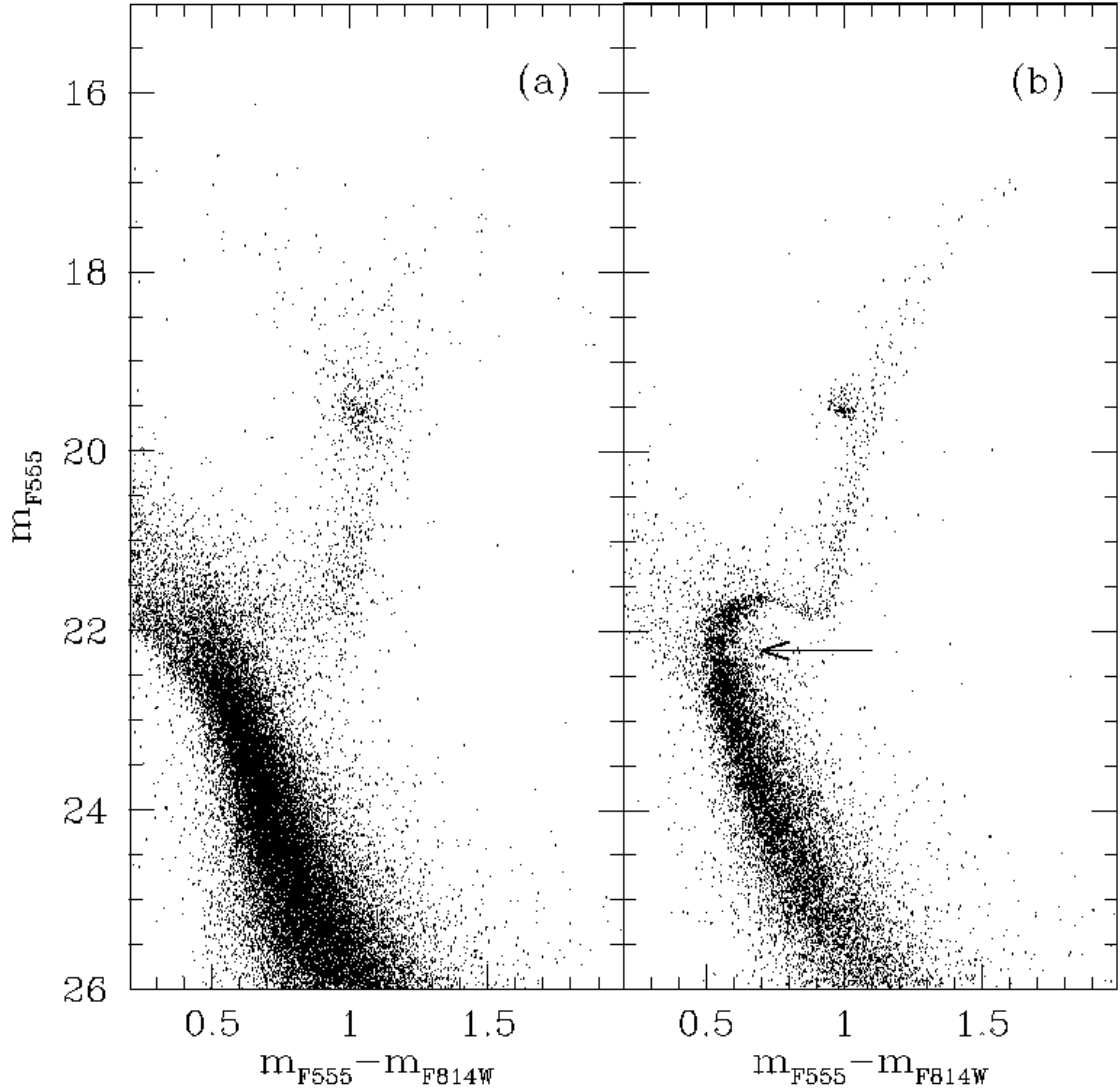}
\caption{\label{f:rgbs} Comparison between the RGB of SMC field~(a), that of
the CMD \#6, shown in Fig~\ref{f:cmds}, corresponding to the intermediate--old
age cluster~(b). The black arrow in panel~(b) highlights the position of the
gap corresponding to the overall contraction of stars, when their central
hydrogen fuel is consumed to a few percent. }
\end{figure}

\begin{figure}
\plotone{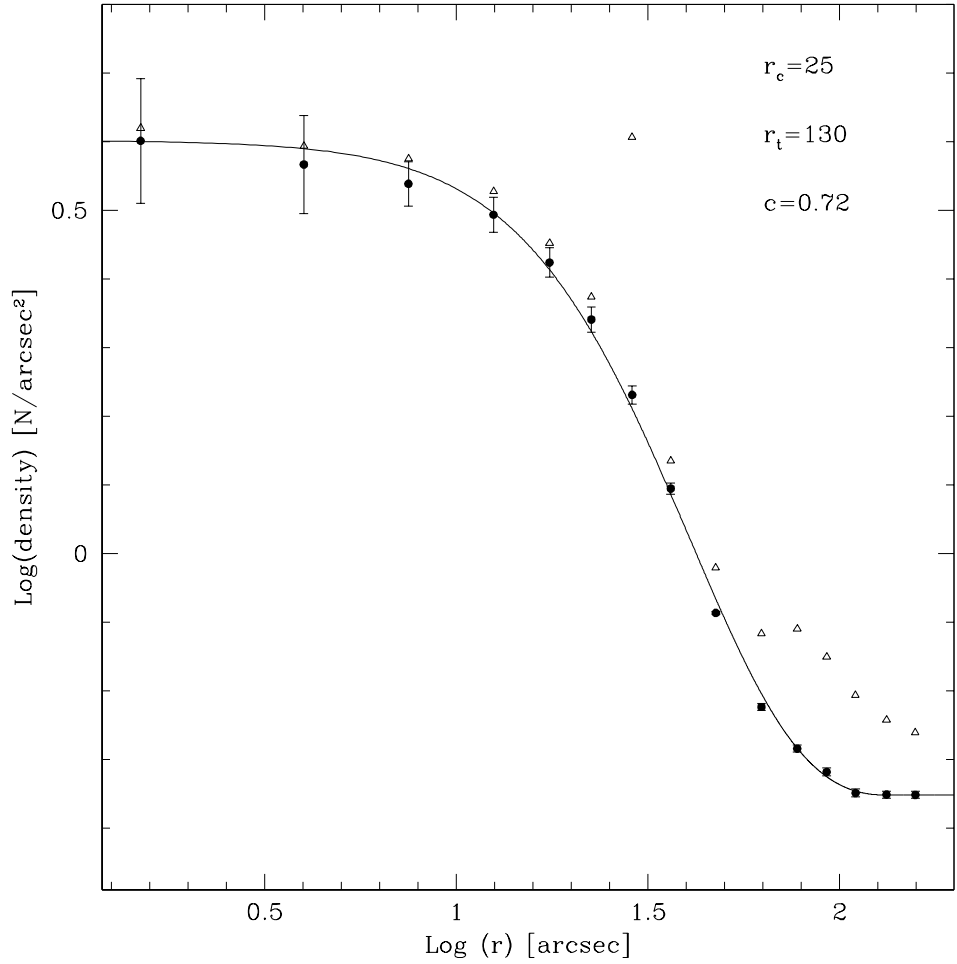}
\caption{\label{f:king} Observed radial density profile of the old cluster
({\it black dots}) and of the entire photometric catalog ({\it open
triangles}), relative to the adopted $C_{grav}$. The solid line indicates the
best-fit model ($r_c=25\arcsec$ and $c=0.72$) of the radial density
distribution of the cluster.}
\end{figure}

\begin{figure}
\plotone{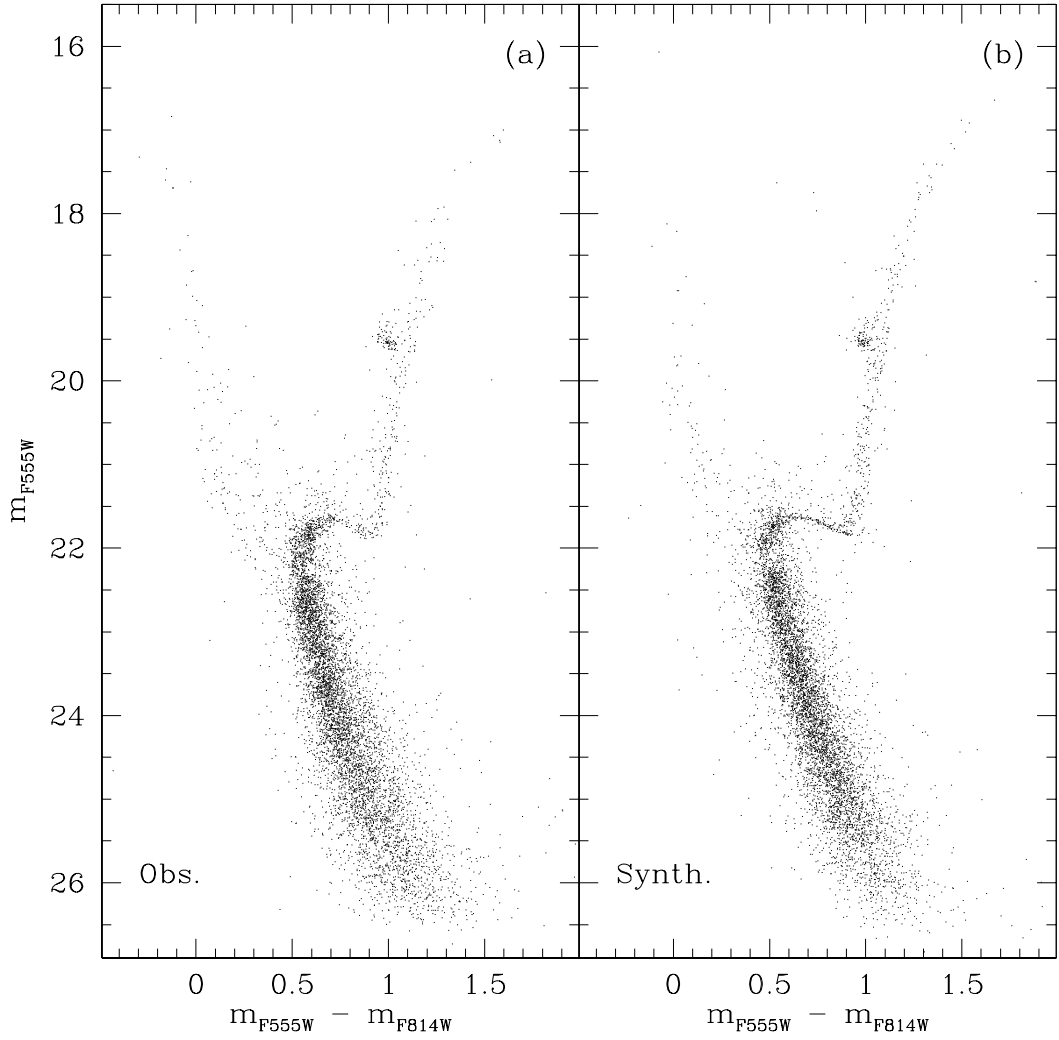}
\caption{\label{f:syn} Comparison between the observed~(a) and the synthetic~(b)
CMDs for the core of BS90. The assumed fraction of binaries is 30\%.} 
\end{figure}

\begin{figure}
\plotone{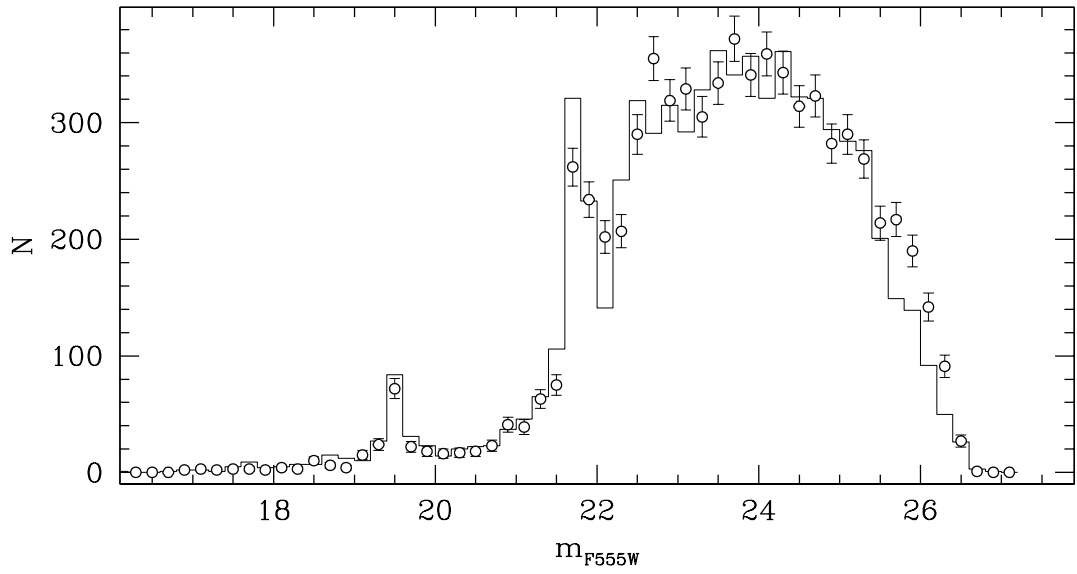}
\caption{\label{f:LFs} Comparison between the observed (open circles) and the
synthetic (solid line) luminosity functions of the core of BS90.}
\end{figure}

An intermediate--old age star cluster is located $\sim$23~pc in
projection from the center of NGC~346. A comparison between the CMDs of this
old cluster and that of the SMC field (see Fig.~\ref{f:rgbs}) indicates that
the stars in the old association formed in an almost instantaneous burst of
star formation $\sim$$4.3\pm 0.1\, {\rm Gyr}$ ago. The discovery of this
intermediate--old age star cluster further supports the hypothesis suggested by
\citet{gardiner92},  \citet{dolphin01}, and \citet{ mccumber05} that, in the
SMC, both the field and the star cluster populations have likely formed in a
quasi--continuous mode.

The youngest stellar population belongs to NGC~346. Padua isochrones
indicate that NGC~346 is $3\pm 1\, {\rm Myr}$ old. The MS of the youngest
population abruptly interrupts at ${\rm F555W}\simeq 21$. At ${\rm F555W}>21$
we identify hundreds of stars whose colors and magnitude are consistent with
those predicted for pre--MS stars (see also Paper~I).

The high spatial resolution of our observations reveals that the youngest stellar
population is not uniformly distributed within the ionized nebula: we identified
at least 15 Sub--clusters, which differ in size and stellar content. Within the
uncertainties due to the comparison with isochrones, the sub--clusters are likely
coeval with each other (see Tab.~\ref{t:sc}). However a relatively older
sub--cluster (Sc--16, with an age $15\pm 2.5\, {\rm Myr}$) is located at the
Northeast periphery of our data. This sub--cluster is likely not related to the
star forming episode that originated NGC~346.

Recent spectroscopic investigations of NGC~346 \citep{heap06,mokiem06} confirm
that the majority of the stars associated with the cluster have an age of 3~Myr, 
even if stars as young as 1~Myr are also present \citep[but see the discussion 
in ][]{massey05}. No relationship between the age of the stars and
their distance from the cluster center is identified. Due to the age
uncertainties quoted by the authors and the lower precision of the  
isochrones comparison technique, our results are in good agreement.

The sub--clusters coincide with the clumps of molecular gas identified by
\citet{rubio00}. Many of them, especially towards the Southeast, are close to
obscure Bok--globules, which can contain still embedded Class~I and 0 young
stellar objects that would not be visible in the wavelength range covered by
our data. The presence of diffuse gas and dust around these sub--clusters
suggests that NGC~346 has likely not exhausted its fuel and star formation is
possibly still ongoing.

All along the dust ``ridge'' of N66 it is possible to observe many dense
nebular knots and dust pillars oriented toward the central clusters, that are
similar in shape to those identified in 30~Doradus by \citet{walborn02}. The
presence of these structures strongly supports the hypothesis that the stellar
winds, developed from the most massive stars in NGC~346, are probably
triggering a secondary episode of star formation at the periphery of N66.

The most massive stars are located in the sub--clusters in the center of the
nebula. We also find that sub--clusters located toward the North--Northeast
are generally smaller and constituted mainly by pre--MS stars. These clusters
coincide with the highest CO velocity clumps analyzed by \citet{rubio00}. This
seems to indicate that the UV radiation and the stellar winds from the stars in
Sc--1, 2, and 3 are not isotropic, but they are more efficient in disrupting
the molecular cloud toward the Northwest, making the star formation less efficient
than toward the Southwest. Alternatively, it is possible that the interaction with SNR
0057--7233 is preventing the normal expansion of N66 along this direction.

NGC~346 is probably the result of the collapse and subsequent fragmentation of
the initial giant molecular cloud in multiple ``seeds'' of star formation. The
fact that the observed sub--clusters appear almost coeval, and that many of
them appear to be connected by arcs of dust and gas, seems to be a good
observational match to the conditions predicted by the hierarchical
fragmentation of a turbulent molecular cloud model
\citep{klessen00,bonnell02,bonnell03}. 

According to this model the fragmentation of the cloud is due to supersonic
turbulent motions present in the gas. The turbulence induces the formation of
shocks in the gas, and produce filamentary structures \citep{bate03}. The
chaotic nature of the turbulence increases locally the density in the
filamentary structures. When regions of high density become self--gravitating,
they start to collapse to form stars. Simulations show that star formation
occur simultaneously at several different location in the cloud
\citep{bonnell03}, as appears to be the case for NGC~346.  

Further UV and IR photometric and spectroscopic observations are necessary to
better establish the mass and the ages of the stars, to verify if
small differences in age are present among the sub--clusters, and to confirm
the presence of ongoing star formation within N66. Furthermore, an analysis of
the gas dynamics within N66 can confirm if the parent cloud underwent a
hierarchical fragmentation: the formation of shocks in the gas, due to the
initial supersonic turbulence, rapidly removes kinetic energy from the gas
\citep{ostriker01}, and we expect to find very low gas velocities around the
sub--clusters.

\acknowledgments
We warmly thank the anonymous referee for the usefull suggestions which helped
us to greatly improve the paper.

We warmly thank Paolo Montegriffo for the software support and Nino Panagia,
Guido De Marchi, and Nolan Walborne for helpful suggestions and useful discussions. The
photometric conversion table to the ACS filters was kindly provided by Livia
Origlia.

The bulk of the synthetic CMD code was originally provided by Laura Greggio. We
thank Giuliana Fiorentino for support on the stellar evolution.

Financial support was provided by the Italian MIUR to ES, LA, and MT, through
Cofin 2002028935 and 2004020323, and STScI GO Grant GO-10248.07-A

\clearpage

\clearpage

\begin{deluxetable}{ccccccc}
\tabletypesize{\small}
\tablewidth{0pt}
\tablecaption{Journal of WFC/ACS observations\label{t:obs}}
\tablehead{
\colhead{Image Name} & \colhead{Filter} & \colhead{Exp.\ Time} & \colhead{R.A.} & \colhead{Dec.} & \colhead{Date} & \colhead{Flag}}
\startdata
J92F01KRQ & ${\rm F555W}$ &   \phn\phn3.0 & $00^{\rm h}59^{\rm m}06\fs72$ & $-72\arcdeg09\arcmin40\farcs3$ & 13 Jul 2004 & N \\
J92F01KSQ & ${\rm F555W}$ & 456.0 & $00^{\rm h}59^{\rm m}06\fs72$ & $-72\arcdeg09\arcmin40\farcs3$ & 13 Jul 2004 & N \\
J92F01KUQ & ${\rm F555W}$ & 456.0 & $00^{\rm h}59^{\rm m}07\fs02$ & $-72\arcdeg09\arcmin40\farcs1$ & 13 Jul 2004 & N \\
J92F01KWQ & ${\rm F555W}$ & 456.0 & $00^{\rm h}59^{\rm m}07\fs31$ & $-72\arcdeg09\arcmin40\farcs0$ & 13 Jul 2004 & N \\
J92F01KYQ & ${\rm F555W}$ & 456.0 & $00^{\rm h}59^{\rm m}07\fs61$ & $-72\arcdeg09\arcmin39\farcs8$ & 13 Jul 2004 & N \\
J92F01L1Q & ${\rm F555W}$ &   \phn\phn3.0 & $00^{\rm h}59^{\rm m}06\fs72$ & $-72\arcdeg09\arcmin40\farcs3$ & 13 Jul 2004 & N \\[0.15cm]
J92FA3VEQ & ${\rm F555W}$ & 380.0 & $00^{\rm h}59^{\rm m}06\fs18$ & $-72\arcdeg10\arcmin31\farcs2$ & 15 Jul 2004 & C \\[0.15cm]
J92F02MBQ & ${\rm F555W}$ &   \phn\phn3.0 & $00^{\rm h}59^{\rm m}04\fs68$ & $-72\arcdeg11\arcmin21\farcs7$ & 18 Jul 2004 & S \\
J92F02MCQ & ${\rm F555W}$ & 483.0 & $00^{\rm h}59^{\rm m}05\fs58$ & $-72\arcdeg11\arcmin21\farcs5$ & 18 Jul 2004 & S \\
J92F02MEQ & ${\rm F555W}$ & 483.0 & $00^{\rm h}59^{\rm m}05\fs28$ & $-72\arcdeg11\arcmin21\farcs6$ & 18 Jul 2004 & S \\
J92F02MGQ & ${\rm F555W}$ & 483.0 & $00^{\rm h}59^{\rm m}04\fs99$ & $-72\arcdeg11\arcmin21\farcs6$ & 18 Jul 2004 & S \\
J92F02MIQ & ${\rm F555W}$ & 483.0 & $00^{\rm h}59^{\rm m}04\fs68$ & $-72\arcdeg11\arcmin21\farcs7$ & 18 Jul 2004 & S \\
J92F02MLQ & ${\rm F555W}$ &   \phn\phn3.0 & $00^{\rm h}59^{\rm m}04\fs68$ & $-72\arcdeg11\arcmin21\farcs7$ & 18 Jul 2004 & S \\[0.15cm]
J92F01L3Q & ${\rm F814W}$ &   \phn\phn2.0 & $00^{\rm h}59^{\rm m}06\fs72$ & $-72\arcdeg09\arcmin40\farcs3$ & 13 Jul 2004 & N \\
J92F01L4Q & ${\rm F814W}$ & 484.0 & $00^{\rm h}59^{\rm m}06\fs72$ & $-72\arcdeg09\arcmin40\farcs3$ & 13 Jul 2004 & N \\
J92F01L6Q & ${\rm F814W}$ & 484.0 & $00^{\rm h}59^{\rm m}07\fs02$ & $-72\arcdeg09\arcmin40\farcs1$ & 13 Jul 2004 & N \\
J92F01L8Q & ${\rm F814W}$ & 484.0 & $00^{\rm h}59^{\rm m}07\fs31$ & $-72\arcdeg09\arcmin40\farcs0$ & 13 Jul 2004 & N \\
J92F01LAQ & ${\rm F814W}$ & 484.0 & $00^{\rm h}59^{\rm m}07\fs61$ & $-72\arcdeg09\arcmin39\farcs8$ & 13 Jul 2004 & N \\
J92F01LDQ & ${\rm F814W}$ &   \phn\phn2.0 & $00^{\rm h}59^{\rm m}06\fs72$ & $-72\arcdeg09\arcmin40\farcs3$ & 13 Jul 2004 & N \\[0.15cm]
J92FA3VGQ & ${\rm F814W}$ & 380.0 & $00^{\rm h}59^{\rm m}06\fs18$ & $-72\arcdeg10\arcmin31\farcs2$ & 15 Jul 2004 & C \\[0.15cm]
J92F02LUQ & ${\rm F814W}$ &   \phn\phn2.0 & $00^{\rm h}59^{\rm m}04\fs68$ & $-72\arcdeg11\arcmin21\farcs7$ & 18 Jul 2004 & S \\
J92F02LVQ & ${\rm F814W}$ & 450.0 & $00^{\rm h}59^{\rm m}05\fs58$ & $-72\arcdeg11\arcmin21\farcs5$ & 18 Jul 2004 & S \\
J92F02LYQ & ${\rm F814W}$ & 450.0 & $00^{\rm h}59^{\rm m}05\fs28$ & $-72\arcdeg11\arcmin21\farcs6$ & 18 Jul 2004 & S \\
J92F02M0Q & ${\rm F814W}$ & 450.0 & $00^{\rm h}59^{\rm m}04\fs99$ & $-72\arcdeg11\arcmin21\farcs6$ & 18 Jul 2004 & S \\
J92F02M5Q & ${\rm F814W}$ & 450.0 & $00^{\rm h}59^{\rm m}04\fs68$ & $-72\arcdeg11\arcmin21\farcs7$ & 18 Jul 2004 & S \\
J92F02M8Q & ${\rm F814W}$ &   \phn\phn2.0 & $00^{\rm h}59^{\rm m}04\fs68$ & $-72\arcdeg11\arcmin21\farcs7$ & 18 Jul 2004 & S 
\enddata
\end{deluxetable}


\begin{deluxetable}{ccccccccc}
\tablewidth{0pt}
\tablecaption{Photometric catalog of the NGC~346 region. RA and Dec are in degrees. The electronic version of the
table contains the entire list. This table is not included in this paper, but it will be made available upon publication.\label{t:cata}}
\tablehead{
\colhead{ID} & \colhead{$m_{F555W}$} & \colhead{$\sigma_{F555W}$} & \colhead{$m_{F814W}$} & \colhead{$\sigma_{F814W}$} &  \colhead{R.A.} & \colhead{Dec.} &\colhead{Reference}}
\startdata
 1 & 11.570 & 0.001 & 11.557 & 0.001 & $0^{\rm h}59^{\rm m}26\fs581$ & $-72\arcdeg09\arcmin53\farcs00$ &	\\
 2 & 11.916 & 0.001 & 11.841 & 0.001 & $0^{\rm h}59^{\rm m}26\fs571$ & $-72\arcdeg09\arcmin53\farcs99$ &	\\
 4 & 12.326 & 0.001 & 11.793 & 0.001 & $0^{\rm h}58^{\rm m}42\fs424$ & $-72\arcdeg09\arcmin43\farcs27$ & MPG185 \\
 5 & 12.551 & 0.002 & 12.397 & 0.001 & $0^{\rm h}59^{\rm m}31\fs975$ & $-72\arcdeg10\arcmin46\farcs11$ & MPG789 \\
 6 & 12.580 & 0.001 & 12.304 & 0.001 & $0^{\rm h}59^{\rm m}26\fs542$ & $-72\arcdeg09\arcmin53\farcs00$ &	\\
 7 & 12.609 & 0.002 & 12.742 & 0.002 & $0^{\rm h}59^{\rm m}04\fs479$ & $-72\arcdeg10\arcmin24\farcs77$ & MPG435 \\
 8 & 12.734 & 0.002 & 12.398 & 0.001 & $0^{\rm h}59^{\rm m}31\fs975$ & $-72\arcdeg10\arcmin46\farcs13$ &	\\
 9 & 13.454 & 0.002 & 13.679 & 0.003 & $0^{\rm h}59^{\rm m}00\fs743$ & $-72\arcdeg10\arcmin28\farcs16$ & MPG355 \\
10 & 13.517 & 0.002 & 11.966 & 0.001 & $0^{\rm h}58^{\rm m}53\fs935$ & $-72\arcdeg12\arcmin04\farcs78$ & MPG283 
\enddata
\end{deluxetable}
 

\begin{deluxetable}{ccccccc}
\tablewidth{0pt}
\tablecaption{Main characteristics of the NGC~346 sub--clusters\label{t:sc}}
\tablehead{
\colhead{Sub--cluster} & \colhead{R.A.} & \colhead{Dec} & \colhead{radius} & \colhead{Age} & \colhead{total } &\colhead{PMS density} \\
                       &  &  & \colhead{parsec} & \colhead{Myr} & \colhead{\# PMS} & \colhead{\# PMS/parsec$^2$}}
\startdata
Sc--1 & $00^{\rm h}59^{\rm m}05\fs2$ & $-72\arcdeg10\arcmin28$ & 2.5 & $3\pm 1$ & \llap{3}42 & 17.4  \\
Sc--2 & $00^{\rm h}59^{\rm m}01\fs8$ & $-72\arcdeg10\arcmin35$ & 2.3 & $3\pm 1$ & \llap{2}95 & 17.8  \\
Sc--3 & $00^{\rm h}59^{\rm m}06\fs0$ & $-72\arcdeg10\arcmin43$ & 1.6 & $3\pm 1$ & \llap{1}38 & 17.2  \\
Sc--4 & $00^{\rm h}59^{\rm m}02\fs5$ & $-72\arcdeg10\arcmin07$ & 0.6 & $3\pm 1$ & 20 & 17.7   \\
Sc--5 & $00^{\rm h}59^{\rm m}00\fs3$ & $-72\arcdeg10\arcmin04$ & 1.1 & $3\pm 1$ & 32 & \phn8.4    \\
Sc--6 & $00^{\rm h}58^{\rm m}57\fs4$ & $-72\arcdeg09\arcmin55$ & 1.2 & $3\pm 1$ & 14 & \phn3.1    \\
Sc--7 & $00^{\rm h}59^{\rm m}05\fs4$ & $-72\arcdeg10\arcmin57$ & 1.5 & $3\pm 1$ & 65 & \phn9.2    \\
Sc--8 & $00^{\rm h}59^{\rm m}07\fs7$ & $-72\arcdeg10\arcmin48$ & 1.9 & $3\pm 1$ & \llap{1}09 & \phn9.6   \\
Sc--9 & $00^{\rm h}59^{\rm m}12\fs5$ & $-72\arcdeg11\arcmin08$ & 1.0 & $3\pm 1$ & 29 & \phn9.2    \\
Sc--10 & $00^{\rm h}59^{\rm m}14\fs9$ & $-72\arcdeg11\arcmin01$ & 1.6 & $3\pm 1$ & 61 & \phn7.6   \\
Sc--11 & $00^{\rm h}59^{\rm m}15\fs3$ & $-72\arcdeg11\arcmin12$ & 1.1 & $3\pm 1$ & 38 & 10.0  \\
Sc--12 & $00^{\rm h}59^{\rm m}19\fs8$ & $-72\arcdeg11\arcmin20$ & 1.5 & $3\pm 1$ & 25 & \phn3.5   \\  
Sc--13 & $00^{\rm h}59^{\rm m}11\fs3$ & $-72\arcdeg10\arcmin00$ & 1.9 & $3\pm 1$ & 90 & \phn7.9   \\
Sc--14 & $00^{\rm h}59^{\rm m}13\fs9$ & $-72\arcdeg09\arcmin27$ & 0.9 & $3\pm 1$ & 22 & \phn8.6   \\
Sc--15 & $00^{\rm h}59^{\rm m}17\fs0$ & $-72\arcdeg09\arcmin15$ & 0.6 & $3\pm 1$ & 21 & 18.6  \\
Sc--16 & $00^{\rm h}59^{\rm m}30\fs8$ & $-72\arcdeg09\arcmin10$ & 1.6 & ~$15\pm 2.5$ & 65 & \phn8.0 
\enddata
\end{deluxetable}


\begin{thebibliography}{}
\bibitem[Angeretti et~al.(2005)]{angeretti05}
Angeretti, L., Tosi, M., Greggio, L., Sabbi, E., Aloisi, A., \& Leitherer, C. 2005,
\aj, 129, 2203
\bibitem[Bate, Bonnell, \& Bromm(2003)]{bate03}
Bate, M.R., Bonnell, I.A., \& Bromm, V. 2003, \mnras, 339, 577
\bibitem[Bica \& Schmitt(1995)]{bica95}
Bica, E.L.D., \& Scmitt, H.R. 1995 \apjs, 101, 41
\bibitem[Bertelli et~al.(1994)]{fagotto94}
Bertelli, G., Bressan, A., Chiosi, C. Fagotto, F., \& Nasi, E. 1994, A\&AS,
106, 275
\bibitem[Bragaglia \& Tosi(2006)]{bragaglia06}
Bragaglia, A., Tosi, M. 2006, \aj, 131, 1544
\bibitem[Bonnell \& Bate(2002)]{bonnell02}
Bonnell, I.A., \& Bate, M.R. 2002, \mnras, 336, 659
\bibitem[Bonnell, Bate, \& Vine(2003)]{bonnell03}
Bonnell, I.A., Bate, M.R., \& Vine, S.G. 2003, \mnras, 343, 413
\bibitem[Bouret et~al.(2003)]{bouret03}
Bouret, J.C., Lanz, T., Hillier, D.J., Heap, S.R., Hubeny, I., Lennon, D. J.,
Smith, L.J., \& Evans, C.J. 2003, \apj, 595, 1182
\bibitem[Calzetti et~al.(1993)]{calzetti93}
Calzetti, D., de Marchi, G., Paresce, F., \& Shara, M. 1993, \apjl, 402, 1
\bibitem[Chabrier(2005)]{chabrier05}
Chabrier, G. 2005, in ``The initial mass function 50 years later'', 41, E.
Corelli, F. Palla, \& H. Zinnecker eds, 2005, ASS library 327, Springer 
\bibitem[Danforth et~al.(2003)]{danfort03}
Danforth, C.W., Sankirt, R., Blair, W.P., Howk, J.C., \& Chu, Y.H. \
2003, \apj, 586, 1179
\bibitem[Dolphin et~al.(2001)]{dolphin01} 
Dolphin, A.E., Walker, A.R., Hodge, P.W., Mateo, M., Olszewski, E.W., Schommer,
R.A. \& Suntzeff, N.B. \ 2001 \apj, 562, 303
\bibitem[Elmegreen(2000)]{elmegreen00}
Elmegreen, B.G. 2000, \aj, 530, 227
\bibitem[Fagotto et~al.(1994a)]{fagotto94a}
Fagotto, F., Bressan, A., Bertelli, G., \& Chiosi, C. 1994a, A\&AS, 104, 365
\bibitem[Fagotto et~al.(1994b)]{fagotto94b}
Fagotto, F., Bressan, A., Bertelli, G., \& Chiosi, C. 1994b, A\&AS, 105, 29
\bibitem[Gardiner \& Hatzidimitriou(1992)]{gardiner92}
Gardiner, L.T., Hatzidimitriou, D. 1992, \mnras, 257, 195
\bibitem[Gilliland(2004)]{gilliland04}
Gilliland. R.L. 2004, Instrument Science Report ACS 04--01, (Baltimore, MD:
STScI)
\bibitem[Haser et~al.(1998)]{haser98}
Haser, S.M., Pauldrach, A.W.A., Leonnon, D.J., Kudrizki, R.--P.,
Lennon, M., Puls, J., \& Voels, S.A. \ 1998, \aap, 330, 285
\bibitem[Heap, Lanz, \& Hubeny(2006)]{heap06}
Heap, S., Lanz, T., \& Hubeny, I. 2006, \apj, 638, 409
\bibitem[Hilditch, Howarth \& Harries(2005)]{hilditch05}
Hilditch, R.W., Howarth, I.D., \& Harries, T.J. 2005, \mnras, 125, 336
\bibitem[Hoopes et~al.(2002)]{hoopes02}
Hoopes, C.G., Sembach, K.R., Howk, J.C., Savage, B.D., \& Fullerton,
A.W. \ 2002, \apj, 569, 233
\bibitem[Izotov, Thuan, \& Lipovetsky(1997)]{izotov97}
Izotov, Y.I., Thuan, T.X., \& Lipovetsky, V.A 1997, \apjs, 108, 1
\bibitem[Izotov \& Thuan(1999)]{izotov99}
Izotov, Y.I., \& Thuan, T.X. 1999, \apj, 511, 639
\bibitem[King(1966)]{king66}
King, I.R. 1966, \aj, 71, 75
\bibitem[Klessen \& Burkert(2000)]{klessen00}
Klessen, R.S., Burkert, A. 2000, \apjs, 128, 287
\bibitem[Koekemoer et~al.(2002)]{koekemoer02}
Koekemoer, A.M., Fruchter, A.S., Hook, R, \& Hack, W. 2002, in {\it Proc 2002
HST Calibration Workshop}, eds. S. Arribas, A. Koekemoer, \& B. Whitmore
(STScI: Baltimore, MD), 337
\bibitem[Koeningsberger, Kuruckz, \& Georgiev(2002)]{kkg02}
Koeningsberger,G.,  Kuruckz, R.L., \& Georgiev, L. \ 2002, \apj,
581, 598
\bibitem[Krist(2003)]{krist03}
Krist, J.E., 2003 ACS Instrument Science Report 2003--06, STScI
\bibitem[Kroupa \& Weidner(2003)]{kroupa03}
Kroupa, P., \& Weidner, C. 2003, \apj, 598, 1076
\bibitem[Massey, Parker, \& Garmany(1989)]{massey89}
Massey, P., Parker, J.W., \& Garmany, C.D. 1989, \aj, 94, 1305
\bibitem[Massey et~al.(2005)]{massey05}
Massey, P., Puls, J., Pauldrach, A.W.A., Bresolin, F., Kudritzki, R.P., Simon, T.
2005, \apj, 627, 477 
\bibitem[McCumber et~al.(2005)]{mccumber05}
McCumber, M.P., Garnett, D.R., Dufour, R.J. 2005, \aj, 130, 108
\bibitem[Mokiem(2006)]{mokiem06}
Mokiem, M.R. 2006, ``The physical properties of early--type massive stars (Ph.D.
Thesys)
\bibitem[Naz\`e et~al.(2002)]{naze02}
Naz\`e, Y., Hartwell, J.M., Stevens, J.R., Corcoran, M.F., Koeningsberger, G.,
Moffat, A.F.J., \& Niemela, V.S. \ 2002, \apj, 580, 225
\bibitem[Naz\`e et~al.(2003)]{naze03}
Naz\`e, Y., Hartwell, J.M., Stevens, J.R.,Manford, J., Marchenko, S., Corcoran,
M.F., Moffat, A.F.J.,\& Skalkowski, G. \ 2003, \apj, 586, 983
\bibitem[Niemela, Marraco, \& Cabanne(1986)]{niemela86}
Niemela, V.S., Marraco, H.G., \& Cabanne, M.L. 1986, \pasp, 98, 1133
\bibitem[Nota et~al.(2006)]{nota06}
Nota, A., Sirianni, M., Sabbi, E., Tosi, M., Meixner, M., Gallagher, J.,
Clampin, M., Oey, S., Smith, L.J., Walterbos, R., \& Mack, J. 2006, \apjl, 640,
29 (Paper~I)
\bibitem[Olive, Skillman, \& Steigman(1997)]{olive97}
Olive, K.A., Skillman, E., \& Steigman, G. 1997, \apj, 483, 788
\bibitem[Origlia \& Leitherer(2000)]{origlia00}
Origlia, L. \& Leitherer, C. 2000, \aj, 119, 2018
\bibitem[Ostriker, Stone, \& Gammie(2001)]{ostriker01}
Ostriker, E.C., Stone, J.M., \& Gammie, C.F. 2001, \apj, 546, 980
\bibitem[Pagel \& Tautvaisiene(1998)]{pagel98}
Pagel, B.E.J, \& Tautvaisiene, G. 1998, \mnras, 299, 535
\bibitem[Pavlovsky et~al.(2004a)]{pavlovsky04a}
Pavlovsky C., et al. 2004, ``ACS Instrument Handbook'', Version 5.0,
(Baltimore: STScI)
\bibitem[Pavlovsky et~al.(2004)]{pavlovsky04}
Pavlovsky, C., Reiss, A., Mack, J., \& Gilliland, R. 2004, ``ACS Data
Handbook'', Version 3.0, (Baltimore: STScI)
\bibitem[Peimbert, Peimbert, \& Ruiz(2000)]{peimbert00}
Peimbert, M., Peimbert, A., \& Ruiz, M.T. 2000, \apj, 541, 688
\bibitem[Peimbert, Peimbert, \& Luridiana(2002)]{peimbert02}
Peimbert, M., Peimbert, A., \& Luridiana V. 2002, \apj, 565, 668
\bibitem[Reid et~al.(2006)]{reid06}
Reid, W. A., Payne, J. L., Filipovi\'c, M. D., Danforth, C. W., Jones, P. A., 
White, G. L., \& Stavely-Smith, L. 2006, \mnras, 367, 1393
\bibitem[Rela\~no, Peimbert, \& Beckman(2002)]{relano02}
Rela\~no, M., Peimbert, M., \& Beckman, J. 2002, \apj, 564, 704
\bibitem[Rich et~al.(2000)]{rich00}
Rich, R.M., Shara, M., Fall, S.M, Zurek, D.2000, \aj, 119, 197
\bibitem[Rubio et al.(2000)]{rubio00}
Rubio, M., Contursi, A., Lequeux, J., Probst, R., Barb\`{a}, R.H., Boulanger,
F., Cesarsky, D., \& Maoli, R. 2000, \aap, 359, 1139
\bibitem[Sabbi, Sirianni \& Nota(2006)]{sabbi06}
Sabbi, E., Sirianni, M., \& Nota, A. 2006, ``The 2005 HST Calibration Workshop, Hubble After the
Transition to Two--Gyro Mode'', ed Koekemoer, Goudfrooij \& Dressel, 41
\bibitem[Siess, Dufour \& Forestini(2000)]{siess00}
Siess, L., Dufour, E., \& Forestini, M. \ 2000, \aap, 358, 593
\bibitem[Sirianni et~al.(2000)]{sirianni00}
Sirianni, M., Nota, A., De Marchi, G., Leitherer, C., \& Clampin, M. 2000, \apj,
579, 288
\bibitem[Sirianni et~al.(2002)]{sirianni02}
Sirianni, M., Nota, A., Leitherer, C., De Marchi, G., \& Clampin, M. 2000, \apj,
533, 203
\bibitem[Sirianni et~al.(2005)]{sirianni05}
Sirianni, M., et al. 2005, \pasp, 117, 1049
\bibitem[Tosi et~al.(1991)]{tosi91}
Tosi, M., Greggio, L., Marconi, G., \& Focardi, P. 1991, \aj, 102, 951
\bibitem[Tosi et~al.(2001)]{tosi01}
Tosi, M., Sabbi, E., Bellazzini, M., Aloisi, A., Greggio, L., Leitherer, C., \&
Montegriffo, P. 2001, \aj, 122, 1271
\bibitem[Walborn(1978)]{walborn78}
Walborn, N.R. \ 1978, \apj, 224L, 133
\bibitem[Walborn \& Blades(1986)]{walborn86}
Walborn, N.R., \& Blades, J.C. 1986, \apjl, 304, 17
\bibitem[Walborn \& Parker(1992)]{walborn92}
Walborn, N.R., \& Parker, J.W. 1992, \apjl, 399, 87
\bibitem[Walborn, Ma\'{i}z--Apell\'{a}niz, \&  Barb\'{a}(2002)]{walborn02}
Walborn, N.R., Ma\'{i}z--Apell\'{a}niz, J.,\&  Barb\'{a}, R.H. 2002, \aj, 124,
1601
\bibitem[Ye, Turtle, \& Kennicutt(1991)]{ye91}
Ye, T., Turtle, A.J., \& Kennicutt, R.C.Jr. 1991, \mnras, 249, 722
\end{thebibliography}
\end{document}